\documentclass[12pt]{article}

\usepackage{scicite}
\usepackage{times}
\usepackage{soul}
\usepackage{url}
\usepackage{comment}
\usepackage[normalem]{ulem}
\useunder{\uline}{\ul}{}
\usepackage{lscape}
\usepackage{amsmath}
\usepackage{amssymb}
\usepackage{booktabs}
\usepackage{graphicx}
\usepackage{upgreek}
\usepackage{textcomp}
\usepackage{float}
\usepackage{pdfpages}
\usepackage[noadjust]{cite}
\newcommand{\RM} [1]{\mathrm{#1}}
\newcommand{\cmpc}{$\RM{cm}^{-3}\,\RM{pc}$}
\newcommand{\radm}{$\mathrm{rad\,m^{-2}}$}
\newcommand{\ergs}{$\mathrm{erg\,s^{-1}}$}
\newcommand{\mhzms}{$\mathrm{MHz\,ms^{-1}}$}
\newcommand{\mhzs}{$\mathrm{MHz\,s^{-1}}$}
\newcommand{\Figure}[1]{Figure~\ref{fig:#1}}
\newcommand{\Table}[1]{Table~\ref{tab:#1}}
\newcommand{\GPM}{GPM\,J1839--10}

\newenvironment{sciabstract}{%
\begin{quote} \bf}
{\end{quote}}

\title{A highly magnetized long-period radio transient exhibiting unusual emission features}

\author
{Yunpeng Men,$^{1\ast\dagger}$
Sam McSweeney,$^{2\dagger}$\\
Natasha Hurley-Walker,$^{2\ast}$
Ewan Barr,$^{1\ast}$
Ben Stappers$^{3}$
\\
\footnotesize{$^{1}$Max-Planck-Institut f{\"u}r Radioastronomie, Auf dem H{\"u}gel 69, Bonn, D-53121, Germany}\\
\footnotesize{$^{2}$International Centre for Radio Astronomy Research, Curtin University, 1 Turner Ave, Bentley, WA, 6102, Australia}\\
\footnotesize{$^{3}$Jodrell Bank Centre for Astrophysics, Department of Physics and Astronomy,}\\
\footnotesize{The University of Manchester, Oxford Road, Manchester, M13 9PL, United Kingdom}\\
\footnotesize{$^\ast$Corresponding author; E-mail: nhw@icrar.org, ebarr@mpifr-bonn.mpg.de, ypmen@mpifr-bonn.mpg.de}\\
\footnotesize{$^\dagger$These authors contributed equally to this work.}
}

\date{}

\begin{document} 


\baselineskip24pt


\maketitle


\begin{sciabstract}
  Long-period radio transients are a new class of astrophysical objects that exhibit periodic radio emission on timescales of tens of minutes. Their true nature remains unknown; possibilities include magnetic white dwarfs, binary systems, or long-period magnetars; the latter class are predicted to produce fast radio bursts (FRBs). Using the MeerKAT radio telescope, we conducted follow-up observations of the long-period radio transient GPM J1839–10.  Here we report the source exhibits a wide range of unusual emission properties, including polarization characteristics indicative of magnetospheric origin, linear-to-circular polarization conversion, and drifting sub-structures closely resembling those observed in repeating FRBs. These radio characteristics provide evidence in support of the long-period magnetar model and suggest a possible connection between long-period radio transients, magnetars, and FRBs.

  Teaser: A long-period radio transient exhibits unusual emission features that could potentially reveal its physical nature.
\end{sciabstract}

% In setting up this template for *Science* papers, we've used both
% the \section* command and the \paragraph* command for topical
% divisions.  Which you use will of course depend on the type of paper
% you're writing.  Review Articles tend to have displayed headings, for
% which \section* is more appropriate; Research Articles, when they have
% formal topical divisions at all, tend to signal them with bold text
% that runs into the paragraph, for which \paragraph* is the right
% choice.  Either way, use the asterisk (*) modifier, as shown, to

\section*{Introduction}

Long-period radio transients (LPRTs), a new identified class of astrophysical objects, exhibit periodic radio emissions on timescales of tens of minutes \cite{Hurley-Walker2022Nat, Hurley-Walker2023Nat, Caleb2024NatAs}. These sources can manifest as either short-lived \cite{Hurley-Walker2022Nat} or long-lived \cite{Hurley-Walker2023Nat} characteristic, and their physical nature remains a subject of active debate \cite{Beniamini2023MNRAS, Qu2024arXiv}. Proposed models include magnetized white dwarfs \cite{Rea2024ApJ}, binary systems \cite{Loeb2022RNAAS}, and long-period magnetars \cite{Tong2023RAA}. Magnetars, characterized as young and highly magnetized neutron stars with higher magnetic field strengths than typical pulsars, ranging from $10^{13}\,\mathrm{G}$ to $10^{15}\,\mathrm{G}$ \cite{Kaspi2017ARA&A}, represent one of the suggested models. Currently, six magnetars are known to exhibit radio emission with periods spanning from 2\,s to 12\,s \cite{Camilo2006Nat, Camilo2007ApJ, Levin2010ApJ, Eatough2013Nat, Champion2020MNRAS, CHIME2020Nat}.

A wide range of similarities exist among the observed emission properties of normal pulsars, millisecond pulsars, and radio magnetars, despite differences in their magnetic field structures and magnetosphere sizes \cite{Philippov2022ARA&A}. These shared emission properties include quasi-periodicity, significant circular polarization, orthogonal polarization modes (OPMs), and the association between the sign change of circular polarization and rapid position angle (PA) swing. The suggestion of the long-period magnetar model proposes that these similar emission properties can also be extended to LPRTs.

Fast radio bursts (FRBs) are bright radio bursts lasting from microseconds to tens of milliseconds \cite{Lorimer2007Sci, CHIME2021ApJS, Snelders2023NatAs}, and their physical origin and emission mechanism remain unknown. However, a Galactic FRB, FRB\,200428, was identified producing from the radio magnetar SGR\,1935+2154 \cite{CHIME2020Nat, Bochenek2020Nat}, suggesting a potential association between magnetars and a subset of FRBs. Therefore, the radio magnetars could explain for both the LPRTs and FRBs. Furthermore, the hypothesis of a long-period magnetar, though not confirmed, has been proposed to elucidate certain FRB properties, including periodic activity \cite{Beniamini2020MNRAS}. The validation of this potential connection remains a subject of ongoing investigation.

{\GPM}, discovered in the Murchison Widefield Array (MWA) observation, has exhibited activity for three decades, featuring a period of $1,318.1957\pm0.0002$\,s \cite{Hurley-Walker2023Nat}. This establishes it as the longest-lived LPRT reported to date. Previous observations revealed that {\GPM} possesses a dispersion measure (DM) of $273.5\pm2.5$\,{\cmpc} and a rotation measure (RM) of $531.63\pm0.15$\,{\radm} \cite{Hurley-Walker2023Nat}. The period derivative constraint of $\dot{P}\lesssim3.6\times10^{-13}\,\mathrm{s}\,\mathrm{s}^{-1}$ was derived from archival data of {\GPM}. The combined characteristics of a slow period and a low period derivative place {\GPM} at the very edge of the most generous death line \cite{Hurley-Walker2023Nat}, which is established based on the rotating dipolar magnetic fields and pair-production mechanisms that typically account for coherent emission from normal pulsars \cite{Zhang2000ApJ}.

\section*{Results and Discussion}
\subsection*{Radio observation and pulse morphology}
A 2.5-hour follow-up polarimetric observation for {\GPM} was conducted with MeerKAT on August 20, 2023 (UT). The observation was performed in the UHF-band, covering frequencies from 544 MHz to 1088 MHz, with a time resolution of 15 $\mu$s. We detected pulsed emissions in three periods, denoted as P1, P2, and P3 in the following text (\Figure{dynamic_spectrum}), which gives a rough event rate of pulsed emission from {\GPM} of 20\%. The dispersion measures (DMs) for the three pulses are $274.5\pm1.9$ {\cmpc}, $274.12\pm0.25$ {\cmpc} and $274.0\pm0.9$ {\cmpc}, which are consistent with the earlier observation on July 20, 2022 (UT) where the DM was measured to be $273.5\pm2.5$ {\cmpc} \cite{Hurley-Walker2023Nat}. The narrowest burst component has a width of about 10\,ms. The pulsed emission reaches a peak flux density ranging between 0.1\,Jy to 0.4\,Jy in the UHF-band, resulting in an isotropic luminosity of $2.1_{-1.5}^{+2.8} \times 10^{30}$ {\ergs} to $8.5_{-6.0}^{+10.8} \times 10^{30}$ {\ergs} (see materials and methods).

These pulses last for about 80 to 120 seconds, exhibiting complex morphologies, despite similarities in the phases with radio emission, accounting for roughly ten percent of the period. The P2 pulse exhibits quasi-periodic emission structures with a quasi-period of 1.97~seconds at a significance level of 3.6-$\sigma$ (see materials and methods). Quasi-periodic emission structures have been identified in normal radio pulsars \cite{Cordes1979AuJPh, Mitra2015ApJ}, rotating radio transients (RRATs) \cite{Chen2022ApJ}, and radio magnetars \cite{Kramer2023NatAs}. An empirical relationship between the rotational period and the period of quasi-periodic structures has been established for these sources \cite{Kramer2023NatAs}. This relationship can be extended to the quasi-periodic structures of the P2 pulse detected in this observation as well (\Figure{quasi-period-relation}).

\subsection*{Down-drifting sub-structure}
With the high time resolution observation of MeerKAT, we discovered that the short-duration sub-structures display drifting behaviors within the dynamic spectrum of {\GPM}. In the P3 pulse, a sub-structure demonstrating multiple emission components displays both single-component down-drifting and multi-component down-drifting characteristics (\Figure{drifting1}). The sub-structure comprises a wideband emission component (C1) along with two narrowband emission components (C2 and C3). The two narrowband emission components can be fitted using a 2D-Gaussian shape (see materials and methods). The emission component C2 exhibits no apparent frequency drifting, featuring a Full Width at Half Maximum (FWHM) bandwidth of $206\pm7$\,MHz and a FWHM width of $4.6\pm0.2$\,ms. Conversely, the emission component C3 displays a FWHM bandwidth of $117\pm4$\,MHz, a FWHM width of $15.2\pm0.5$\,ms, and a single-component down-drifting rate of $-5.3\pm0.5$\,{\mhzms}. There is a central time offset of $17.4\pm0.2$\,ms between the two emission components C2 and C3. Their respective center frequencies are $920\pm3$\,MHz and $630\pm2$\,MHz, showcasing a multi-component down-drifting rate of $-16.6\pm0.2$\,{\mhzms}.

Frequency drifting behavior has been observed in FRBs \cite{CHIME2019Nat, CHIME2019ApJ, Hessels2019ApJ, Day2020MNRAS, Fonseca2020ApJ} and solar bursts \cite{Pick2008A&ARv}. The frequency drifting behavior observed in solar bursts typically involves longer timescales and slower frequency drifting rates compared to this sub-structure of {\GPM}. For instance, solar type III radio bursts usually exhibit a higher frequency drifting rate, around 0.1\,{\mhzms}, and a duration of a few seconds \cite{Reid2014RAA}. In contrast, repeating FRBs showcase diverse features, including multi-component down-drifting (also known as the `sad-trombone effect') and single-component down-drifting \cite{Platts2021MNRAS, Zhou2022RAA, Jahns2023MNRAS}. The frequency drifting rates in these events can vary from a few {\mhzms} to several thousand {\mhzms} \cite{Hessels2019ApJ, CHIME2019ApJ, Fonseca2020ApJ, Hilmarsson2021MNRAS, Zhou2022RAA, Jahns2023MNRAS}. The drifting rate observed in this sub-structure of {\GPM} is akin to the drifting rates witnessed in CHIME repeating bursts observed within a similar frequency band \cite{CHIME2019ApJ,Fonseca2020ApJ}. Additionally, bursts from repeating FRBs exhibit narrowband emission features, with bandwidths ranging from approximately 50\,MHz to 400\,MHz within the 400-800\,MHz band \cite{Pleunis2021}. These durations of these bursts typically span from a few milliseconds to tens of milliseconds \cite{Pleunis2021}. Both the bandwidth and duration of the emission components in this sub-structure of {\GPM} fall within the range observed in CHIME repeating FRBs (\Figure{down-drifting-compare}). This similarity in frequency drifting behavior suggests a shared emission mechanism or propagation effect between {\GPM} and repeating FRBs \cite{Wang2019ApJ, Lyutikov2020ApJ, Metzger2022ApJ, Kundu2021MNRAS}.

\subsection*{Polarization properties}
\begin{figure*}[h!]
    \centering
    \includegraphics[width=0.8\textwidth]{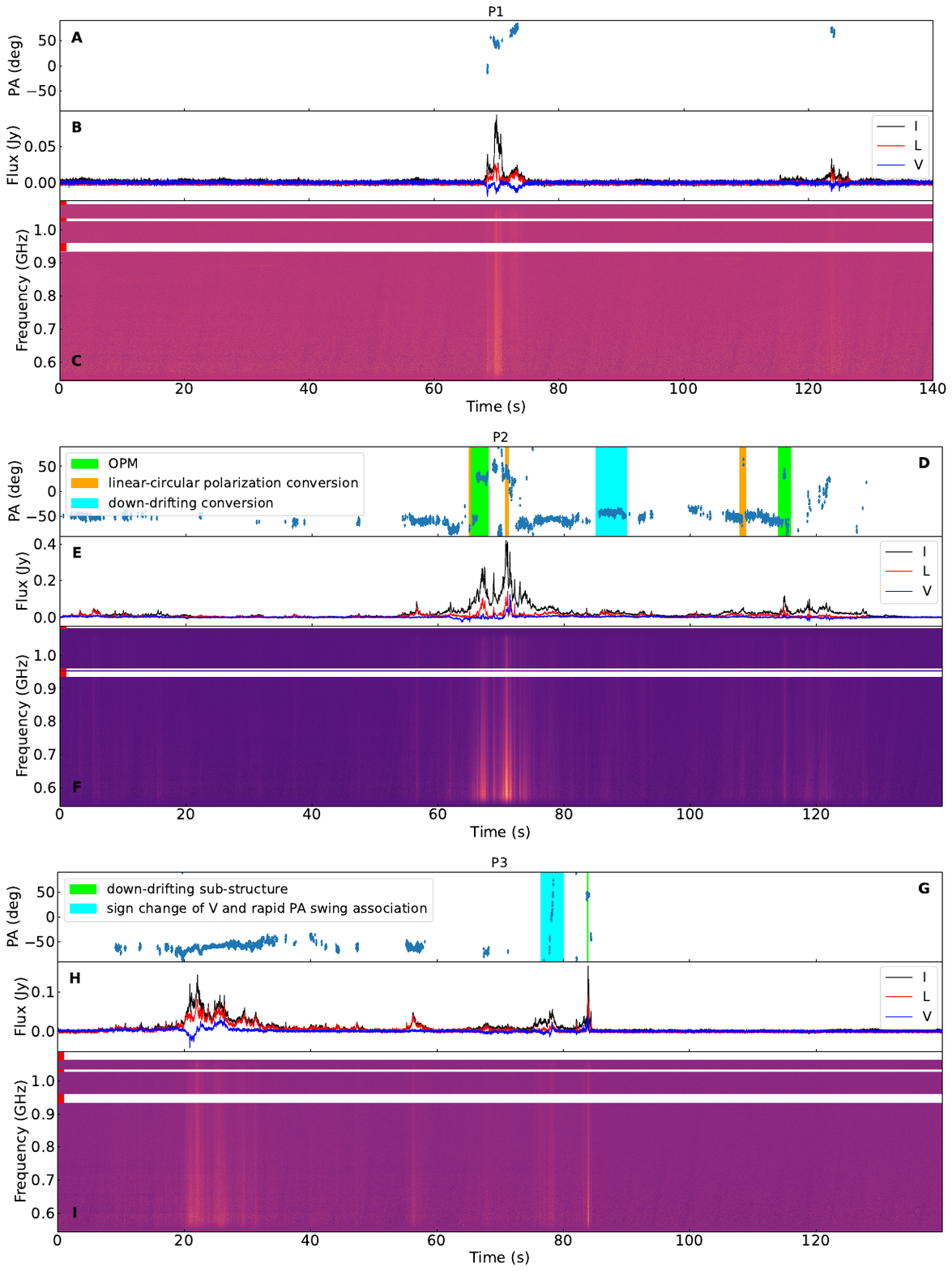}
    \caption{\textbf{Dynamic spectra aligned in phase of the three pulses detected on 20 August 2023.} (A) (D) (G) show the time evolution of the position angle (PA). (B) (E) (H) show the temporal variations of total intensity (I, black), linear polarization intensity (L, red), and circular polarization intensity (V, blue). (C) (F) (I) show the dynamic spectrum of total intensity corrected for dispersion at $274.12 \pm 0.25$\,{\cmpc}. The Faraday correction was applied using RMs of $531.05 \pm 0.02$\,{\radm}, $530.76 \pm 0.02$\,{\radm}, and $530.76 \pm 0.01$\,{\radm} for each pulse, respectively. The frequency channels affected by RFI contamination are mitigated and indicated by white strips on the dynamic spectrum (see materials and methods).}
    \label{fig:dynamic_spectrum}
\end{figure*}

\begin{figure*}[h!]
    \centering
    \includegraphics[width=0.8\textwidth]{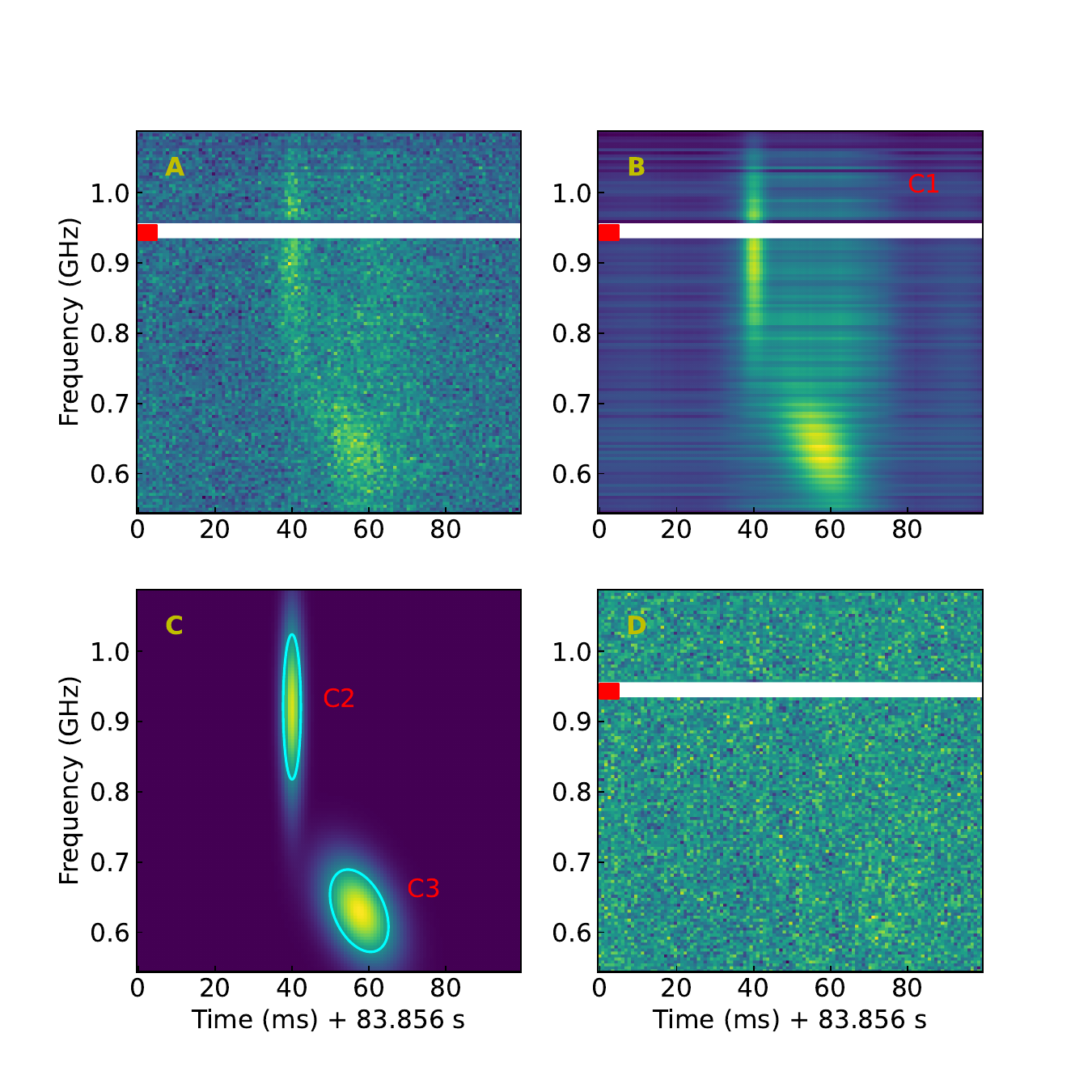}
    \caption{\textbf{Spectrum fitting of the down-drifting sub-structure within the P3 pulse.} (A) Dynamic spectrum. (B) Fitted dynamic spectrum showcasing the broadband emission C1 with the two narrowband emission components C2 and C3. (C) Fitted dynamic spectrum depicting the emission components C2 and C3. The cyan elliptical circle denotes the FWHM width of the fitted 2D-Gaussian shape (see materials and methods). (D) Residual dynamic spectrum. The time is aligned with the P3 pulse in \Figure{dynamic_spectrum}.}
    \label{fig:drifting1}
\end{figure*}

Using the MeerKAT full-Stokes data, we measured the rotation measures (RMs) of the three pulses, which are $530.68\pm0.05$\,{\radm}, $530.33\pm0.05$\,{\radm}, and $530.44\pm0.04$\,{\radm}, respectively (see materials and methods). The RM variation among these three pulses is less than 1\,{\radm}. Additionally, the RM change from the previous observation on July 20, 2022 (UT) is also less than 2\,{\radm}, with an RM value of $531.63\pm0.15$\,{\radm} (\Figure{rm_dm_change}). Upon correction for Faraday rotation using the measured RMs, the polarized profile of the three pulses was derived. The polarization characteristics exhibit complex variations across the observed period. The linear polarization (LP) fraction shows fluctuations ranging from approximately 0\% to 100\% across different phases, whereas the circular polarization (CP) fraction reaches values as high as about 95\%, exhibiting an association between the sign change and rapid PA swing at specific phases (\Figure{sign_change}). Furthermore, the PA swing exhibits complex variations with phase, being relatively constant in some phase ranges while displaying steep changes in others (\Figure{dynamic_spectrum}), and OPMs are detected in the PA swing (\Figure{opm}).

The PA swing and occurrence of OPMs are frequently observed in the polarization emission of pulsars. The PA swing is attributed to variation in the direction of the magnetic field as the line-of-sight crosses the emission region \cite{Everett2001ApJ}. Additionally, the association between the rapid PA swing and the sign change of CP was suggested to be caused by the geometry effect when the line of sight sweeps the emission cones in a pulsar magnetosphere \cite{Radhakrishnan1990ApJ}. Furthermore, while the PA swing in pulsars is often explained by the rotating vector model (RVM) \cite{Everett2001ApJ}, this model does not apply straightforwardly to {\GPM} , as it exhibits multiple irregular PA changes within approximately one-tenth of its period (\Figure{dynamic_spectrum}). This discrepancy indicates that the emission region of {\GPM} possess a complex magnetic field configuration similar to radio magnetars \cite{Kramer2007MNRAS, Dai2019ApJ, Tong2021MNRAS}. The PA swings can be explained by magnetospheric models as opposed to the synchrotron maser mechanism because the latter predicts a constant PA profile for the emitted radiation due to the requirement of ordered magnetic fields in the shocked
region where the emission takes place. The observed PA swing profiles in {\GPM}, therefore, favor a magnetospheric origin. It should be noted that diverse PA swings were also observed for FRBs, and a similar interpretation was drawn as well \cite{Luo2020Nat}. Moreover, the emission of {\GPM} exhibits significant circular polarization and linear polarization fractions in certain phases, akin to the polarization properties observed in normal pulsars \cite{Dai2015MNRAS}, radio magnetars \cite{Dai2019ApJ}, and FRBs \cite{Xu2022Nat, Jiang2022RAA}. This similarity implies the presence of common emission mechanisms or propagation effects among them.

\subsection*{Linear-to-circular polarization conversion}
\begin{figure*}[h!]
    \centering
    \includegraphics[width=0.8\textwidth]{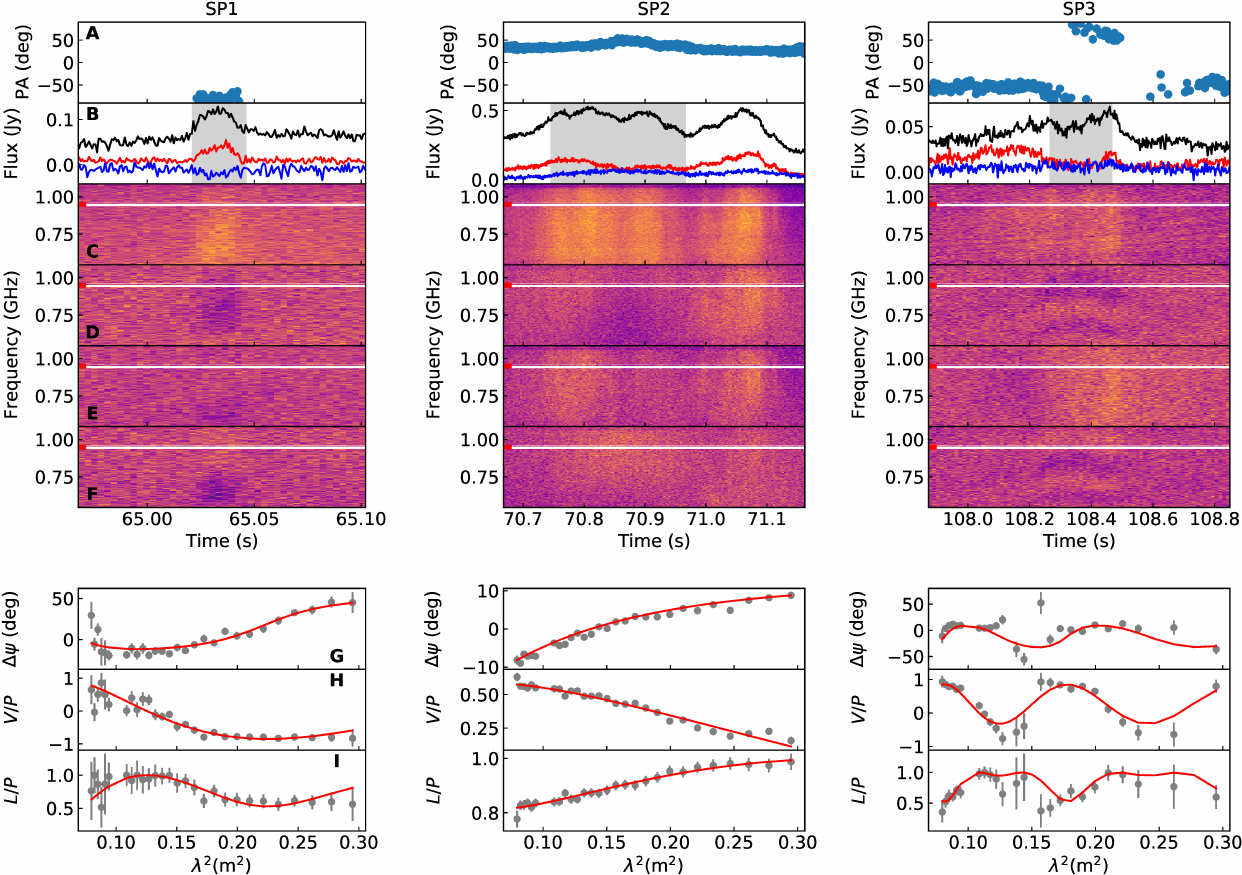}
    \caption{\textbf{Stokes spectra of the sub-pulses, exhibiting linear-to-circular polarization conversion within the P2 pulse.} (A) PA variation. (B) Total intensity ($I$, black), linear polarization intensity ($L$, red) and circular polarization intensity ($V$, blue). (C) Stokes $I$ spectrum. (D) Stokes $Q$ spectrum. (E) Stokes $U$ spectrum. (F) Stokes $V$ spectrum. (G) Faraday conversion fitting of the PA residuals after the Faraday rotation correction (see materials and methods). (H) Faraday conversion fitting of the normalized circular polarization residuals. (I) Faraday conversion fitting of the normalized linear polarization residuals.}
    \label{fig:faraday_conversion}
\end{figure*}

Within the sub-pulses of P2, there are evident instances of frequency-dependent conversion between linear and circular polarized emission (\Figure{faraday_conversion}). To investigate this behavior, we focused on three sub-pulses (referred to as SP1, SP2, and SP3 hereafter) within P2, demonstrating considerable frequency-dependent polarization variation. Notably, the polarization behavior varies across different phases of the pulse. Therefore, we extracted the PA residuals, linear and circular polarization fractions within the specified phase range exhibiting stable frequency-dependent polarization variations (see materials and methods). SP1, characterized by its short duration, exhibits a sign change in the CP  spectrum. Moreover, both the LP fraction and PA residuals showcase frequency-dependent variations. SP2 falls within the phase range marked by the peak flux of P2, demonstrating a noticeable decrease in CP fraction and an increase in LP fraction. Similarly, its PAs exhibit frequency-dependent residuals. In SP3, oscillations are observed in both CP and LP, alongside rapid variations in PA concerning the pulse phase. To characterize the polarization variation, we utilized a phenomenological model \cite{Lower2021arXiv} to fit the spectrum (see materials and methods). The generalized rotation measures (GRMs) of SP1, SP2 and SP3 are $20.01^{+0.45}_{-8.8}$, $4.16^{+0.44}_{-0.49}$, and $45.97^{+1.0}_{-0.6}$ with corresponding wavelength-dependent indices of $0.35^{+0.39}_{-0.24}$, $0.84^{+0.23}_{-0.16}$, and $0.95^{+0.17}_{-0.13}$.

The linear-to-circular polarization conversion was also observed in FRB\,20201124A \cite{Xu2022Nat} and the radio-loud magnetar XTE\,J1810-197 \cite{Lower2023arXiv}. FRB\,20201124A was reported to exhibit linear-to-circular polarization conversion with wavelength-dependent indices of 2 \cite{Xu2022Nat} and 3 \cite{Kumar2023PhRvD}. In contrast, the magnetar XTE\,J1810-197, which shows linear-to-circular polarization conversion with a wavelength-dependent index ranging from 0 to 3 \cite{Lower2023arXiv}, is more similar to {\GPM}. Furthermore, the limited RM variation indicates the near-field origin of its polarization conversion behavior, which can be produced in the magnetosphere. These findings support the presence of a similar local magneto-ionic environment around {\GPM} and XTE\,J1810-197. Given that Faraday conversion typically occurs in strong magnetic fields \cite{Gruzinov2019ApJ}, we can potentially estimate the local magnetic field strength where the polarization conversions take place, $B\sim\frac{300}{\gamma}\,\mathrm{G}$ (see Supplementary Text).

\subsection*{Down-drifting polarization conversion}
In the P2 pulse, we observed a statistically significant down-drifting polarization conversion behavior in the Stokes spectra (\Figure{drift_conversion}). A distinct down-drifting band exhibited prominent linear-to-circular polarization conversion. Outside this band, the emission displayed a small CP fraction, while within the band, CP emission intensified. Moreover, the total intensity decreased by an average of about 50\% in the down-drifting band. The drift rate was $4.2_{-0.1}^{+0.1}$\,{\mhzs} with a frequency relation index of $1.95_{-0.61}^{+0.03}$ (see materials and methods). This radio emission characteristic has not been reported previously. The current theoretical models for down-drifting structures \cite{Wang2019ApJ, Lyutikov2020ApJ, Metzger2022ApJ, Kundu2021MNRAS} and polarization conversion behaviors \cite{Gruzinov2019ApJ} face challenges in explaining the observed concurrence of the two emission characteristics. This presents an unprecedented opportunity to investigate both down-drifting and polarization conversion mechanisms.

\subsection*{Implications}
The physical nature of LPRTs remains mysterious \cite{Hurley-Walker2022Nat, Hurley-Walker2023Nat}, with various models proposed, such as magnetized white dwarfs \cite{Rea2024ApJ}, binary systems \cite{Loeb2022RNAAS}, and long-period magnetars \cite{Tong2023RAA}. In this study, we present the discovery of diverse emission properties in {\GPM} that resemble those observed in radio magnetars or pulsars, include quasi-periodicity, orthogonal polarization modes, significant circular and linear polarization, and the association between the sign change of circular polarization and rapid PA swing, which are produced from their magnetospheres. These observed similar radio emission properties provide evidence for the magnetospheric origin of the radio emission of {\GPM}. Furthermore, {\GPM} exhibits a similar behavior of linear-to-circular polarization conversion as the magnetar XTE\,J1810--197, which indicates a similar magnetosphere environment around {\GPM} with radio magnetars, supporting the long-period magnetar model. Additionally, a down-drifting sub-structure observed in {\GPM} resembles the repeating FRBs. The observed similarities in emission properties in our observation suggest a possible connection between LPRTs, magnetars, and FRBs.

\section*{Materials and Methods}
\subsection*{Radio observations}
MeerKAT, operated by the South African Radio Astronomy Observatory, is a large radio interferometer comprising 64 dishes, each 13.5 meters in size, situated in the Northern Cape of South Africa \cite{Jonas2016mks}. The follow-up observation of {\GPM} was conducted under the project code DDT-20220718-NH-01 in the UHF-band, spanning a frequency range of 544 to 1088 MHz. The observations comprised 15 sessions, each lasting approximately 10 minutes, equivalent to half of the period of {\GPM}. This observation scheme allows for the complete coverage of 15 periods within a total observation time of 2.5 hours (\Table{observation}). The full polarization data were recorded using the PTUSE backend \cite{Bailes2020PASA}, operating in the PSRFITS search mode, employing 1024 frequency channels, and a time resolution of 15\,$\mu$s. Notably, the recorded data has already undergone polarization calibration \cite{Serylak2021MNRAS}. We processed the data using {\textsc{DSPSR}} \cite{Straten2011PASA} to generate archive files containing the full polarization spectrum with a time resolution of 12\,ms post-downsampling. Consequently, only last three periods exhibited detectable radio emission. For the remaining periods, we conducted a single pulse search using {\textsc{TransientX}} \cite{Men2024arXiv}, employing a pulse width search range below 100\,ms, a dispersion measure (DM) range of 250-300\,{\cmpc}, and a signal-to-noise ratio (S/N) threshold of 8. No radio pulses were detected in this search. Subsequently, we conducted DM estimation for the three identified pulses using {\textsc{DM\_phase}} \cite{Seymour2019ascl}. This method extracts DMs by maximizing the pulse structures. The DMs obtained for the three pulses are $274.5 \pm 1.9$\,{\cmpc}, $274.12 \pm 0.25$\,{\cmpc}, and $274.0 \pm 0.9$\,{\cmpc}, respectively. The rotation measures (RMs) for the three pulses were obtained by fitting the Stokes Q and U spectrum (\Figure{qu_fitting}), with a detailed description of the algorithm available in \cite{Luo2020Nat}. The derived RMs for the three pulses are $531.05 \pm 0.02$\,{\radm}, $530.67 \pm 0.02$\,{\radm}, and $530.76 \pm 0.01$\,{\radm}, respectively. It should be noted that we corrected the sign of the RM for the observation on July 20, 2022 \cite{Hurley-Walker2023Nat} because the receiver handedness was not corrected. To account for the ionosphere's RM contribution, the code {\textsc{ionFR}} \cite{Sotomayor2013A&A} was applied. This correction resulted in revised RMs of $530.68 \pm 0.05$\,{\radm}, $530.33 \pm 0.05$\,{\radm}, and $530.44 \pm 0.04$\,{\radm} for the three pulses. Notably, there were no significant changes observed in either DM or RM compared to the observation conducted on July 20, 2022 (\Figure{rm_dm_change}).

\subsection*{Radio luminosity}
The flux density of the pulses were estimated using the radiometer equation,
\begin{equation}
    S = \mathrm{S/N} \times \frac{T_\mathrm{sys}}{G \, \sqrt{n_p \, \mathrm{BW} \, T_\mathrm{s}}}\,,
\end{equation}
where the signal-to-noise ratio (S/N) is estimated by the ratio of the intensity of a sample to noise. The system temperature $T_\mathrm{sys}$ is about 18\,K and the total antenna gain $G$ is about 2.8\,K/Jy \cite{Bailes2020PASA}. $n_p$ is 2, which is the number of polarization summed. The bandwidth BW is 544\,MHz, while the time resolution is 12\,ms. To estimate the luminosity, we use the DM distance of $5.7 \pm 2.9$\,kpc estimated with the YMW16 model \cite{Yao2017ApJ, Hurley-Walker2023Nat}. The derived peak isotropic luminosities $L=4\pi\,d_\mathrm{DM}^2\,S$ are $2.1_{-1.5}^{+2.8} \times 10^{30}$\,erg/s, $8.5_{-6.0}^{+10.8} \times 10^{30}$\,erg/s and $3.5_{-2.7}^{+4.5} \times 10^{30}$\,erg/s.

\subsection*{Quasi-periodicity}
The P2 pulse exhibits multiple peak structures in it's profile. To investigate the quasi-periodicity, we conducted the periodicity analysis using the auto-correlation function (ACF) and power spectral density (PSD). The ACF was computed by calculating the autocorrelation coefficient ($\rho$) of the total intensity ($I$) within the time range of 60-80\,s, determined as
\begin{equation}
    \rho(n\,T_s) = \frac{\sum_i I[t_i] \cdot I[t_i+n\,T_s]}{\sum_i I[t_i] \cdot I[t_i]}\,,
\end{equation}
where $T_s$ is the time resolution of the intensity profile, and the time lag is $n\,T_s$. The PSD was derived from the same total intensity series using the \textsc{scipy.signal.periodogram}.
A quasi-period of 1.97\,s is revealed in the ACF and PSD plots (\Figure{quasi-periodicity}). We choose  the quasi-period 1.97\,s rather than the double period based on the intensity profile peaks. To assess the significance of this quasi-periodicity, we applied Fisher's g-test based on the PSD \cite{Nowroozi1967}. As the test is conducted at the maximum PSD peak, the upper bound of the p-value for the quasi-period was estimated of $p<2.7\times10^{-4}$, equivalent to 3.6 sigma. We roughly measured the error of the quasi-period based on the width of the quasi-periodic structures $\sim 1$\,s, which gives a geometric standard deviation (GSTD) factor of 1.5 \cite{Kramer2023NatAs}. We extracted the data in the work \cite{Kramer2023NatAs}, and added the rotational period and period of quasi-periodic structures {\GPM} to the relation, which can be fit in 2-sigma (\Figure{quasi-period-relation}).

\subsection*{Down-drifting structure}
In the pulse of the period P3, we found a sub-structure exhibiting frequency down-drifting behavior (\Figure{drifting1}). To investigate the drifting sub-structure, we segmented a 100-millisecond data segment around it, employing a time resolution of 0.2\,ms using {\textsc{DSPSR}}. We performed de-dispersion and Faraday rotation correction using the DM of 274\,{\cmpc} and RM of $530.76$\,{\radm}. To mitigate radio frequency interference (RFI), we manually identified and removed contaminated frequency channels using {\textsc{pazi}} in {\textsc{PSRCHIVE}}. The data was then normalized to achieve unit variance and a zero offset in each frequency channel for subsequent analysis, using the off-pulse data. The sub-structure comprises three emission components: a broadband emission component (C1) with a wide profile and two narrowband components of shorter durations in the upper (C2) and lower bands (C3) respectively. To investigate the down-drifting structure, we characterized the total intensity spectrum $I$ using a model given by

\begin{align*}
    I_{j,i} &= I_{0, j, i} + I_{1, j, i} + I_{2, j, i}\,, \\
    I_{0, j, i} &= c_j s_i + d_j\,, \\
    I_{1, j, i} &= f(x_i, y_j, K_1, x_{c, 1}, y_{c, 1}, a_1, b_1, \theta_1)\,, \\
    I_{2, j, i} &= f(x_i, y_j, K_2, x_{c, 2}, y_{c, 2}, a_2, b_2, \theta_2)\,, \\
    x_i &= \frac{t_i}{0.1\,\mathrm{s}}\,,\\
    y_i &= \frac{f_i}{1\,\mathrm{GHz}}\,,
\end{align*}
where $I_0$, $I_1$, and $I_2$ represent the total intensity of emission components C1, C2, and C3. $c_j$ and $d_j$ stand for the amplitude and baseline offset of the common profile template $s$ in the $j$-th frequency channel of C1, respectively. Additionally, $t_i$ and $f_j$ denote time and frequency, respectively. The function $f$ represents a 2D-Gaussian shape,
\begin{align*}
    f(x, y, K, x_c, y_c, a, b, \theta) = K \exp \bigg[
    &-\left(\frac{\cos^2{\theta}}{a^2} + \frac{\sin^2{\theta}}{b^2}\right) (x - x_c)^2 \\
    &-2 \left(\frac{\sin{\theta} \cos{\theta}}{a^2}
    + \frac{\sin{\theta} \cos{\theta}}{b^2}\right) (x - x_c) (y - y_c) \\
    &-\left(\frac{\sin^2{\theta}}{a^2} + \frac{\cos^2{\theta}}{b^2}\right) (y - y_c)^2 \bigg]\,, \\
\end{align*}
where $K$ denotes the peak intensity, and $\theta$ represents the declination angle. $x_c$ and $y_c$ denote the rescaled central time and frequency, respectively. $a$ and $b$ signify the rescaled duration and bandwidth. The FWHM bandwidth $B_\mathrm{FWHM}$ and duration $W_\mathrm{FWHM}$ of the emission component using the 2D-Gaussian shape parameters are expressed as
\begin{align*}
    B_\mathrm{FWHM} &= 2 \sqrt{\log2} B \times 1\,\mathrm{GHz}\,,\\
    W_\mathrm{FWHM} &= 2 \sqrt{\log2} W \times 0.1\,\mathrm{s}\,,\\
    B &= \sqrt{\frac{4 p}{4 p q - r^2}}\,,\\
    W &= \sqrt{\frac{4 q}{4 p q - r^2}}\,,\\
    p &= \frac{\cos^2{\theta}}{a^2} + \frac{\sin^2{\theta}}{b^2}\,,\\
    q &= \frac{\sin^2{\theta}}{a^2} + \frac{\cos^2{\theta}}{b^2}\,,\\
    r &= 2 \sin{\theta} \cos{\theta} \left( \frac{1}{b^2} - \frac{1}{a^2} \right)\,,
\end{align*}
which represent the marginalized width along the time and frequency dimension, respectively. The single-component down-drifting rate $D_\mathrm{s}$ is
\begin{equation}
    D_\mathrm{s} = \frac{\cos{\theta}}{\sin{\theta}} \times 10\,\mathrm{MHz\,s^{-1}}\,,
\end{equation}
and the multi-component down-drifting rate $D_\mathrm{m}$ is calculated by dividing the center frequency and time offset of the emission component C2 and C3,
\begin{equation}
    D_\mathrm{m} = \frac{y_{c,2}-y_{c,1}}{x_{c,2}-x_{c,1}} \times 10\,\mathrm{MHz\,s^{-1}}\,.
\end{equation}
To estimate the profile $s$ of the emission component C1, we extract the average profile from the spectrum within the frequency range of 720-730\,MHz. This range is selected to minimize the signal contribution from emission components C2 and C3. The obtained average profile was subsequently smoothed using the Savitzky-Golay filter \cite{Savitzky1964AnaCh} to create the template $s$. To estimate the parameters within the total intensity spectrum model, we employed the sampler {\textsc{Multinest}} \cite{Feroz2008MNRAS}, which relies on the nested sampling algorithm \cite{Skilling2004AIPC}. Uniform priors were assigned to all parameters, and the resultant posterior distributions are depicted in \Figure{drifting_posteriors}. The emission component C2 exhibits a derived FWHM bandwidth of $206\pm7$\,MHz and a duration of $4.6\pm0.2$\,ms, while the emission component C3 displays a FWHM bandwidth of $117\pm4$\,MHz and a duration of $15.2\pm0.5$\,ms. The single-component down-drifting rate of emission component C3 is measured at $-5.3\pm0.5$\,{\mhzms}, whereas the multi-component down-drifting rate observed for C2 and C3 stands at $-16.6\pm0.2$\,{\mhzms}.

\subsection*{Linear-to-circular polarization conversion}
Within the P2 pulse, multiple sub-pulses display noticeable frequency-dependent variations in the full-Stokes spectrum. Among them, one sub-pulse demonstrates oscillations in polarization that vary with frequency. To investigate this polarization behavior, we focused on the data around three sub-pulses (SP1, SP2, SP3) for further analysis. The data was initially normalized using the off-pulse range data, and we manually masked the frequency channels contaminated by RFI signals. The dispersion and Faraday rotation were corrected with the DM of 274.12\,{\cmpc} and RM of 530.67\,{\radm} estimated using the entire pulse of P2. We derived the frequency-dependent position angle (PA) residual $\Delta \psi_j$, from the Stokes Q and U spectra, while correcting the time-dependent PA based on the equation
\begin{align*}
    \widehat{Q}_j &= \sum_i \left[ Q_{j,i} \cos (2 \psi_i) + U_{j,i} \sin (2\psi_i) \right]\,,\\
    \widehat{U}_j &= \sum_i \left[ U_{j,i} \cos (2 \psi_i) - Q_{j,i} \sin (2\psi_i) \right]\,,\\
    \widehat{L}_j &= \sqrt{\widehat{Q}_j^2 + \widehat{U}_j^2}\,,\\
    \widehat{P}_j &= \sqrt{\widehat{Q}_j^2 + \widehat{U}_j^2 + \widehat{V}_j^2}\,,\\
    \Delta \psi_j &= \frac{1}{2}\arctan \frac{\widehat{U}_j}{\widehat{Q}_j} - \mathrm{RM}\,\lambda_j^2 \,,
\end{align*}
where $\psi_i$ represents the PA of the $i$-th time sample, while $\lambda_j$ signifies the wavelength of the $j$-th frequency channel. Additionally, we derived the frequency-dependent linear and circular polarization fractions, incorporating the generalized Weisberg correction in \cite{Everett2001ApJ, Jiang2022RAA}. These processes are conducted solely within time ranges, showing consistent and stable residuals across all three sub-pulses due to the varying nature of the polarization characteristics over time. Moreover, we enhanced the estimation of the PA residual and polarization fraction by aggregating values from neighboring frequency channels. The residual Stokes spectrum is depicted in \Figure{faraday_conversion} and the polarization vector variation on Poincaré sphere is shown in \Figure{poincare}, illustrating the linear-to-circular polarization conversion. Notably, SP3 exhibits frequency-dependent oscillations in the polarization fraction and PA residual. To characterize this polarization behavior, we applied the phenomenological model proposed in \cite{Lower2023arXiv}. In our study, we employed a distinct representation, formulated as
\begin{equation}
    \pmb{p}_j = \pmb{R}(\pmb{u}, \phi_j) \cdot \pmb{p_0} \,,
\end{equation}
where $\pmb{p}_j$ represents the observed unit polarization vector normalized by the total polarization intensity,
\begin{align*}
    \pmb{p}_j &= \frac{1}{P_j}\begin{bmatrix}
           Q_j \\
           U_j \\
           V_j
         \end{bmatrix}\,.
\end{align*}
$\pmb{R}(\pmb{u},\phi_j)$ denotes the 3D-rotation matrix with the polarization rotation axis $\pmb{u}$ and the rotation angle $\phi_j$ in the $j$-th frequency channel. The matrix $\pmb{R}(\pmb{u},\phi)$ is defined as
\begin{multline*}
    \pmb{R}(\pmb{u},\phi) = \\
    \begin{bmatrix}
    \cos{\phi}+u_x^2 (1-\cos{\phi}) & u_x u_y (1-\cos{\phi})-u_z \sin{\phi} & u_x u_z (1-\cos{\phi})+u_y \sin{\phi} \\
    u_y u_x (1-\cos{\phi})+u_z \sin{\phi} & \cos{\phi}+u_y^2 (1-\cos{\phi}) & u_y u_z (1-\cos{\phi})-u_x \sin{\phi} \\
    u_z u_x (1-\cos{\phi})-u_y \sin{\phi} & u_z u_y (1-\cos{\phi})+u_x \sin{\phi} & \cos{\phi}+u_z^2 (1-\cos{\phi})
    \end{bmatrix}\,.\\
\end{multline*}
The rotation angle $\phi_j$ is a function of wavelength, which is expressed as
\begin{equation}
    \phi_j = \mathrm{GRM} \left(\left(\frac{\lambda}{1\,\mathrm{m}}\right)^\alpha-\left(\frac{\lambda_c}{1\,\mathrm{m}}\right)^\alpha\right)\,\mathrm{rad}\,,
\end{equation}
where GRM is the `generalized' rotation measure and $\alpha$ is the wavelength-dependent index. $\lambda_c$ is the corresponding wavelength at central observation frequency of 784\,MHz. We utilized the {\sc{MULTINEST}} sampler to derive the posterior distributions and parameter estimations for fitting the frequency-dependent polarization vector variation. The resulting GRMs for SP1, SP2, and SP3 are $20.01^{+0.45}_{-8.8}$, $4.16^{+0.44}_{-0.49}$, and $45.97^{+1.0}_{-0.6}$, respectively. The corresponding wavelength-dependent indices are $0.35^{+0.39}_{-0.24}$, $0.84^{+0.23}_{-0.16}$, and $0.95^{+0.17}_{-0.13}$. It has been proposed that generalized Faraday rotation could result in phase-resolved variations of RM \cite{Lower2023arXiv}. To explore this possibility, we conducted RM fittings for each 2-second data block. Notably, the RMs exhibit significant variations across the pulse phase in P2, particularly during the time interval characterized by peak flux and rapid PA variations (\Figure{RM_variation2}). The RM variations should not be attributed to instrument artifacts, as the RM for some pulsars observed with the same instruments remains stable within $<1$\,{\radm} over the timescale of days \cite{Keith2024MNRAS}. However, the possibility that these variations arise from instantaneous changes in magnetic fields along the line of sight can't be excluded.

\subsection*{Down-drifting polarization conversion}
\begin{figure*}[h!]
    \centering
    \includegraphics[width=0.7\textwidth]{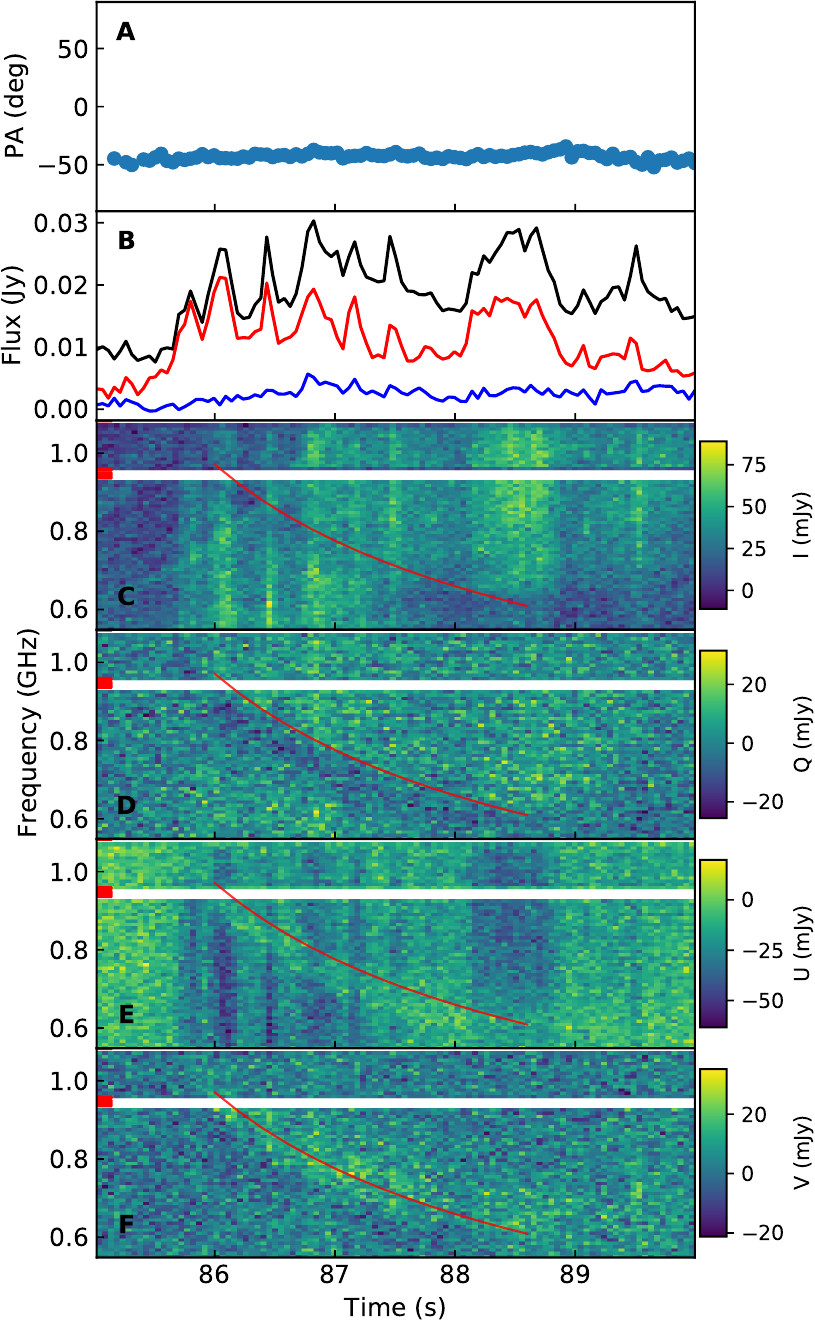}
    \caption{\textbf{Stokes spectra of the sub-structure with down-drifting polarization conversion within the P2 pulse.} (A) PA variation. (B) Total intensity ($I$, black), linear polarization intensity ($L$, red) and circular polarization intensity ($V$, blue). (C) Stokes $I$ spectrum. (D) Stokes $Q$ spectrum. (E) Stokes $U$ spectrum. (F) Stokes $V$ spectrum. The red lines in the spectra represent the fitted frequency-dependent relation of the down-drifting conversion (see materials and methods).}
    \label{fig:drift_conversion}
\end{figure*}

In the P2 pulse, a down-drifting linear-to-circular polarization conversion structure was observed in the Stokes spectrum. This down-drifting conversion cannot be attributed to polarization calibration issues, as the system is expected to remain stable at such short timescales. Moreover, it is challenging to produce a drifting feature through calibration issues. To further investigate this behavior, we selected the full-polarization data within the time span of 85-90\,s in P2. The Faraday rotation was corrected with the RM of $530.67$\,{\radm}. The resulting Stokes spectrum are shown in \Figure{drift_conversion}, where a down-drifting band with linear-to-circular polarization conversion can be observed. To characterize this behavior, we created residual plots of the frequency-dependent PA and CP/LP fraction for successive 0.2 s segments (\Figure{drift_conversion_fitting}). The PAs in segments exhibit pulsed increases with the central time drifting along the frequency. The CP fractions in segments show drifting pulsed increases, while the LP fractions show drifting pulsed decreases. Additionally, the total intensity shows drifting pulsed decreases. To quantify the down-drifting conversion, we fitted the frequency-dependent PA and CP/LP fraction variations using Gaussian shapes, where the central frequency $\nu$ of the pulsed variation is modeled as
\begin{equation}
    \frac{\nu}{1\,\mathrm{GHz}} = \left(\frac{t - t_c}{D/241\,{\rm s}} + \left(\frac{\nu_c}{1\,\mathrm{GHz}}\right)^{-\beta}\right)^{-1/\beta}\,,
\end{equation}
where $t$ is the central time of a segment, $t_c$ is the mean time of the entire time span, and $\nu_c$ is the corresponding central frequency of the pulsed variation at $t_c$. $\beta$ is the frequency-dependent relation index of the drifting. The down-drifting relation model is similar to the dispersion relation with $\beta=2$, where $D$ is the value of the `equivalent' dispersion measure. For frequency-dependent PA variations and LP/CP variations, we used different central frequencies $v_c$. The resulting $D$ is $390_{-55}^{+217}$ with the corresponding $\beta=1.95_{-0.61}^{+0.03}$. To characterize the total intensity decreases, we modeled the total intensity variations using Gaussian templates with the same time-frequency drifting relation obtained in the LP/CP fitting. The average decrease in total intensity is approximately 50\%.

\clearpage

\clearpage

\clearpage

\clearpage

\clearpage

% Your references go at the end of the main text, and before the
% figures.  For this document we've used BibTeX, the .bib file
% scibib.bib, and the .bst file Science.bst.  The package scicite.sty
% was included to format the reference numbers according to *Science*
% style.

%BibTeX users: After compilation, comment out the following two lines and paste in
% the generated .bbl file. 

\bibliography{scibib}

\bibliographystyle{ScienceAdvances}

\section*{Acknowledgments}
\paragraph{Acknowledgments}
The MeerKAT telescope is operated by the South African Radio Astronomy Observatory, which is a facility of the National Research Foundation, an agency of the Department of Science and Innovation. SARAO acknowledges the ongoing advice and calibration of GPS systems by the National Metrology Institute of South Africa (NMISA) and the time space reference systems department of the Paris Observatory. 
YPM and SJM acknowledge discussion with M. E. Lower. YPM acknowledge discussion or comments from G. Desvignes, M. Kramer, Y. H. Qu, L. Spitler and K. J. Lee.

\paragraph{Funding}
YPM and EB acknowledge continuing support from the Max Planck society. NH-W is supported by an Australian Research Council Future Fellowship (project number FT190100231) funded by the Australian Government.

\paragraph{Author Contributions}
NH-W and EB led the radio proposals. NH-W, EB and BS coordinated the observation. YPM modeled the data and drafted the manuscript with suggestions from all authors. YPM discovered the drifting pulses and down-drifting conversion. SJM discovered the Faraday conversion behavior.

\paragraph{Competing interests}
The authors declare no competing interests.

\paragraph{Data and materials availability} All the archive files used in the analysis in this work will be made publicly available upon acceptance on Dryad (https://doi.org/10.5061/dryad.ghx3ffbww).

%Here you should list the contents of your Supplementary Materials -- below is an example. 
%You should include a list of Supplementary figures, Tables, and any references that appear only in the SM. 
%Note that the reference numbering continues from the main text to the SM.
% In the example below, Refs. 4-10 were cited only in the SM.     
\section*{Supplementary materials}
%Materials and Methods\\
Supplementary Text\\
Figs. S1 to S11\\
Tables S1\\
%References \textit{(47-65)}

% For your review copy (i.e., the file you initially send in for
% evaluation), you can use the {figure} environment and the
% \includegraphics command to stream your figures into the text, placing
% all figures at the end.  For the final, revised manuscript for
% acceptance and production, however, PostScript or other graphics
% should not be streamed into your compliled file.  Instead, set
% captions as simple paragraphs (with a \noindent tag), setting them
% off from the rest of the text with a \clearpage as shown  below, and
% submit figures as separate files according to the Art Department's
% instructions.
\clearpage
\renewcommand{\thesection}{S\arabic{section}}
\renewcommand{\thetable}{S\arabic{table}}
\renewcommand\thefigure{S\arabic{figure}}
\renewcommand{\theequation}{S\arabic{equation}}
\pagenumbering{arabic} 
\setcounter{figure}{0}

\begin{center}
\Huge Supplementary materials for:\\
\Large A highly magnetized long-period radio transient exhibiting unusual emission features

\vspace{6mm}
\author
{\normalsize
{Yunpeng Men,
Sam McSweeney,
Natasha Hurley-Walker,
Ewan Barr,
Ben Stappers}
\\
\vspace{3mm}
\normalsize Corresponding author: E-mail: nhw@icrar.org, ebarr@mpifr-bonn.mpg.de, ypmen@mpifr-bonn.mpg.de
}
\end{center}

\baselineskip24pt

\vspace{2cm}
\Large This PDF includes:\\
%\normalsize Materials and Methods\\
\normalsize Supplementary Text\\
\normalsize Figs. S1 to S11\\
\normalsize Tables S1\\
%\normalsize References \textit{(47-65)}

\textwidth 16cm 
\textheight 21cm
\footskip 1.0cm% 
\baselineskip 24pt

\clearpage

\section*{Supplementary Text}
\subsection*{Theoretical interpretation}
The observations of {\GPM} exhibits a wide range of radio emission properties: (a) Orthogonal polarization modes (OPMs), where smooth position angle (PA) swings are interrupted by approximately 90-degree transitions; (b) Significant circular polarization, sometimes reaching tens of percent; (c) Association of the PA of linear polarization with the sign change of circular polarization; (d) Linear-to-circular polarization conversion; (e) Down-drifting polarization conversion, where the linear polarization converts to circular polarization in a down-drifting frequency band. Among these properties, OPMs, significant circular polarization, and its association with rapid PA changes are common in normal pulsars, millisecond pulsars, and radio magnetars, despite differences in magnetic field structures and magnetosphere sizes \cite{Philippov2022ARA&A}. These radio emission characteristics can be attributed to either intrinsic emission mechanisms or propagation effects within the magnetosphere, or external influences such as a magneto-ionic environment in a binary system or a supernova remnant. Below, we explore possible interpretations of these radio properties in various scenarios.

{\bf{Orthogonal polarization modes}} In the propagation of natural waves through the ultrarelativistic, highly magnetized plasma in a pulsar magnetosphere, only three normal wave modes are allowed \cite{Petrova2001A&A, Philippov2022ARA&A}: the extraordinary X-mode, where the electric field vector is perpendicular to both the wave vector and the magnetic field; the superluminal mode; and the subluminal Alfvén mode, with the electric field in the plane of the wave vector and the magnetic field. The Alfvén mode experiences Landau damping for oblique propagation along magnetic field lines, so the observed orthogonal polarization modes (OPMs) consist of the extraordinary X-mode and the superluminal O-mode. The occurrence of OPMs can be explained by propagation effects in the magnetosphere: (a)  Cyclotron absorption or induced scattering occasionally occurs for one of the modes \cite{Manchester1975ApJ} (b) The two modes are separated in space and angle due to different refractive indices \cite{Melrose1979AuJPh}; (c) The superluminal O-mode transforms into the extraordinary X-mode \cite{Petrova2001A&A}.

{\bf{Circular polarization and its association with linear polarization}} 
Significant circular polarization is observed in the radio emission of {\GPM}. This circular polarization can be produced by intrinsic emission mechanisms within the magnetosphere or by propagation effects inside or outside the magnetosphere \cite{Qu2023MNRAS}. The coherent radio emission in the magnetosphere is proposed to arise from coherent curvature radiation by charged bunches with high Lorentz factors or from coherent inverse Compton scattering (ICS). In the curvature radiation scenario, circular polarization can be produced when the line-of-sight is off the curved magnetic field plane, which can also result in an antisymmetric profile of circular polarization associated with a rapid position angle change \cite{Radhakrishnan1990ApJ, Qu2023MNRAS}. However, the circular polarized radio emission of {\GPM} is also observed without an antisymmetric profile or associated linear polarization, indicating the need for additional mechanisms to account for the observed circular polarization. The coherent ICS process can occur in the magnetosphere when low-frequency electromagnetic waves produced by inner gap sparking are scattered by charged bunches \cite{Qiao1998A&A, Qu2024ApJ}. This scattering can generate circular polarization in an off-beam geometry because the electric fields of the scattered waves can have different phases due to the spatial distribution of electrons within the charged bunches \cite{Qu2023MNRAS}. Given that {\GPM} has an ultra-long period, the Goldreich-Julian density is very low, significantly reducing the total number of net charges in one bunch. Therefore, a twisted magnetic field might be required rather than a normal dipole magnetic field, as proposed to revise the death line of long-period radio transients \cite{Tong2023RAA}.

Circular polarization can also be generated through propagation effects. Within the magnetosphere, cyclotron absorption between the left and right circular polarized radio emissions can differ, producing net circular polarization, which requires the electrons and positrons to have an asymmetric energy distribution. Outside the magnetosphere, it has been noted that synchrotron maser and absorption processes struggle to generate a high degree of circular polarization \cite{Qu2023MNRAS}. Cyclotron absorption, on the other hand, is an efficient mechanism for producing high degrees of circular polarization in dense magneto-ionic environments \cite{Qu2023MNRAS}. For cyclotron absorption to occur, the magnetic field strength in the magneto-ionic region should be $B \approx \gamma (1-\beta \cos\theta) \frac{2 \pi \nu m_e c} {e}$, where $\theta$ is the angle between the wave's incident direction and the electron's motion direction. Assuming the radio waves propagate parallel to the magnetic field, $\theta \approx 0^\circ$, the magnetic field strength $B \approx \frac{180\,\mathrm{G}}{\gamma} \frac{\nu}{1\,\mathrm{GHz}}$. For non-parallel incident waves, such as $\theta=30^\circ$, the magnetic field strength could be underestimated, yielding $B = 48\,\gamma\,\mathrm{G}$, corresponding to a smaller magnetosphere radius. Optical observations suggest the possible presence of a main sequence star with a spectral type ranging from mid-K to mid-M \cite{Hurley-Walker2023Nat}, which disfavors highly magnetized companions such as Be-stars or O-stars \cite{Qu2023MNRAS}. Therefore, relativistic plasma in the magneto-ionic environment is required for cyclotron absorption in such a binary system. However, This condition can be satisfied within the magnetosphere of a neutron star, where the Lorentz factor $\gamma > 100$, and the magnetic field strength is $B \approx 4\times10^{-7} \left(\frac{B_\mathrm{surf}}{10^{14}\,\mathrm{G}}\right)\,\left(\frac{r}{r_\mathrm{c}}\right)^{-3}\,\mathrm{G}$, with the light-cylinder radius of {\GPM} is $r_\mathrm{c} = \frac{P c}{2 \pi} \approx 6.3 \times 10^{7}\,\mathrm{km}$. Under these constraints, the propagation effect emerged at a radius of $< 0.005\,r_\mathrm{c}$, resulting in a small pulse duty cycle in a dipole magnetic field, which is inconsistent with observations. This indicates that the magnetic field configuration of {\GPM} might be more complex. Another mechanism for generating circular polarization is the Faraday conversion effect. Faraday conversion within the magnetosphere is suggested to produce symmetric profiles of circular polarization \cite{Radhakrishnan1990ApJ}, which may account for some of the circular polarized radio emissions observed in {\GPM}.

{\bf{Linear-to-circular polarization conversion}} Linear-to-circular polarization conversion occurs when radio waves propagating through a birefringent medium, manifesting as elliptically or linearly polarized natural wave modes \cite{Kennett1998PASA, Gruzinov2019ApJ}, which is called Faraday conversion. The existence of magnetic field perpendicular to the wave direction can result in elliptically natural wave modes \cite{Kennett1998PASA, Gruzinov2019ApJ}. The wavelength-dependent indices can vary under different plasma conditions, for instance, $\alpha=3$ in the relativistic pair-plasma \cite{Kennett1998PASA}, and $\alpha=1-2$ in the near-wind of a magnetar \cite{Lyutikov2022ApJ}. In the scenario of radio waves propagating through a birefringent magneto-ionetic medium, Faraday conversion becomes significant when the Larmor frequency $\nu_B$ is comparable to the radio frequency $\nu/\gamma$, where the required magnetic field strength is $B\approx \frac{360\,\mathrm{G}}{\gamma} \frac{\nu}{1\,\mathrm{GHz}}$ \cite{Gruzinov2019ApJ, Xu2022Nat, Vedantham2019MNRAS}. This provides a rough estimate of the magnetic field strength in the region where the conversions take place, $\gamma B\sim300\,\mathrm{G}$, which aligns with the estimate from cyclotron absorption. However, it is important to note that cyclotron absorption occurs when radio waves propagate parallel to the magnetic field, while Faraday conversion occurs when they propagate perpendicular to the magnetic field. These propagation effects are therefore expected to occur in distinct regions.

Linear-to-circular polarization conversion can arise from the radio wave propagation effects in the magnetosphere \cite{Cheng1979ApJ}, which can also explain its phased-related variation observed in {\GPM}. It has been suggested that linear-to-circular polarization conversion can occur in the near-wind of a magnetar with a wavelength-dependent index of 1-2 \cite{Lyutikov2022ApJ}, which can be applied to the observed behavior of {\GPM}. As the calculations in the previous section, the conditions for a magneto-ionic medium can be satisfied within the magnetosphere. Polarization conversion can also occur when radio emission passes through the solar wind of a highly magnetized companion, such as a Be-star or a supernova remnant \cite{Qu2023MNRAS}. However, optical observations disfavor the scenario of a highly magnetized companion \cite{Hurley-Walker2023Nat}. In both cases, the magnetic field strength should not exceed 100\,G, implying that the presence of relativistic plasma is required for the observed effects.

The linear-to-circular polarization conversion could also be generated from the multi-path scattering effects when radio waves pass through a magnetized screen \cite{Beniamini2022MNRAS}. The required condition of RM induced by the screen, $\mathrm{RM_s} \gtrsim 10 \left(\frac{\nu}{1\,\mathrm{GHz}}\right)^2\,\mathrm{rad\,m^{-2}}$, can be satisfied for {\GPM}. The variability timescale can be estimated as, $t_\mathrm{scr,var} > (6.7\,\mathrm{s})\nu_\mathrm{co,100Hz}^{1/2} \nu_\mathrm{1GHz}^{-1} d_\mathrm{5.7kpc}^{1/2}v_\mathrm{max,7}^{-1}$, where the correlation bandwidth $\nu_\mathrm{co}$ should exceed 100\,Hz, given the shortest pulse width is less than 10\,ms, because scattering tails are not observed in the short-time structures. This scenario is difficult to explain the various frequency-dependent relationships in the polarization conversion in short time scales. The timescale associated with the magnetized screen makes it difficult to account for the rapid PA changes occurring within hundreds of milliseconds. However, this scenario remains applicable during phases with circularly polarized emission and nearly constant PAs.

{\bf{Down-drifting polarization conversion}}
The spectra of {\GPM} exhibited an unusual emission feature of down-drifting polarization conversion. Within a drifting sub-band, the emission converted from linear polarization to circular polarization, while outside this sub-band, no conversion occurred. Additionally, within the sub-band, the total intensity decreased by an average of about 50\%. This could be evidence of cyclotron absorption, where left and right circular polarization were asymmetrically absorbed. The drifting in the absorption frequencies could be attributed to different magnetic field strengths in the absorption region. In this scenario, the magnetic field strength can be estimated using the cyclotron frequency $B=\frac{360\,\mathrm{G}}{\gamma} \frac{\nu}{1\,\mathrm{GHz}}$ \cite{Qu2023MNRAS}.

\clearpage

\begin{figure}[h!] 
    \centering
    \includegraphics[width=\textwidth]{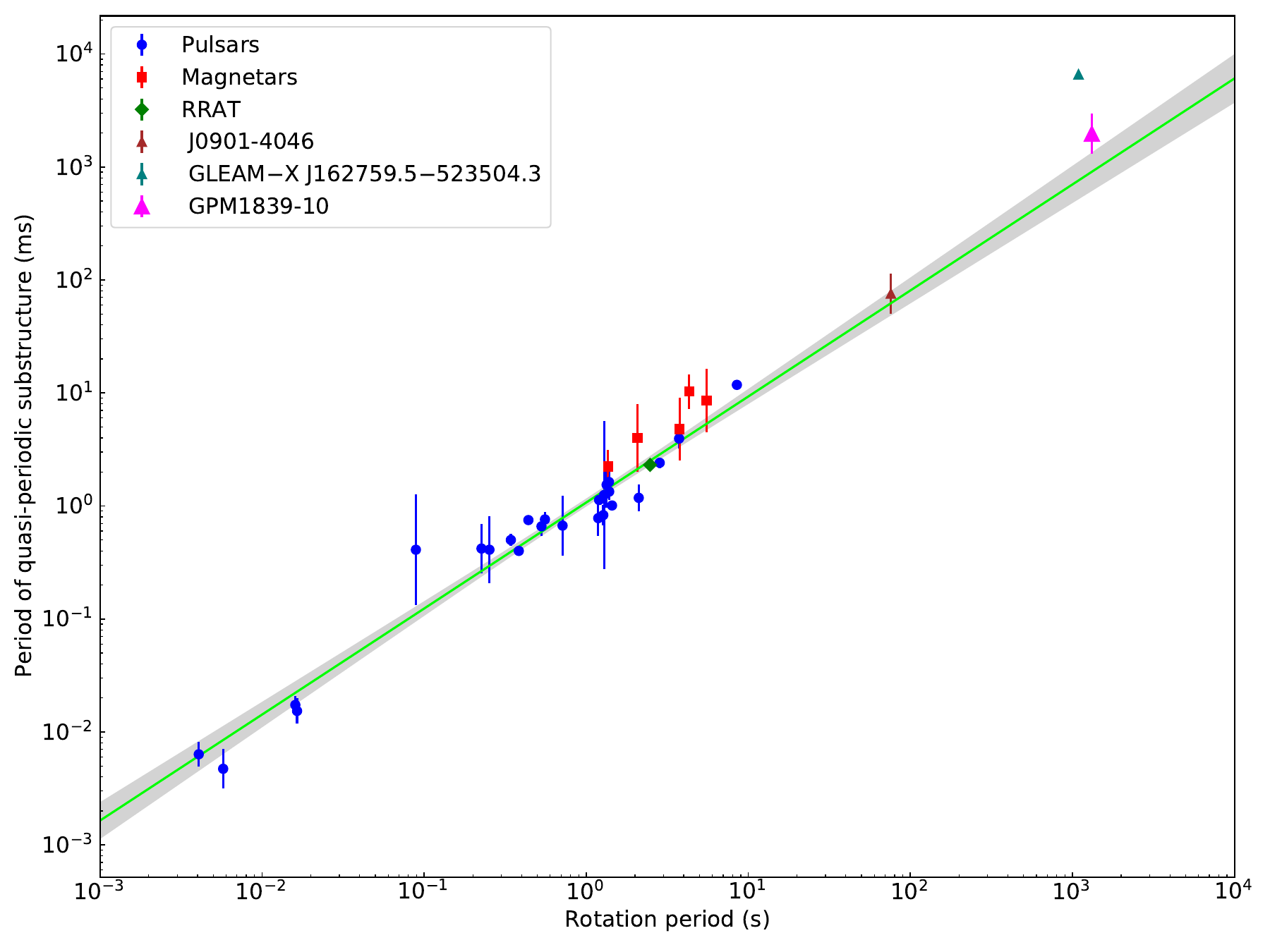}
    \caption{\textbf{Relationship between quasi-periodicity and the observed sub-structure as a function of the neutron star rotation period.} It was delineated in the reference by \cite{Kramer2023NatAs}. The parameters of {\GPM} observed in the P2 pulse are denoted by the pink upper triangle. We re-estimated the 1-sigma confidence interval using the maximum likelihood estimation method.}
    \label{fig:quasi-period-relation}
\end{figure}

\clearpage

\begin{figure}[h!]
    \centering
    \includegraphics[width=\textwidth]{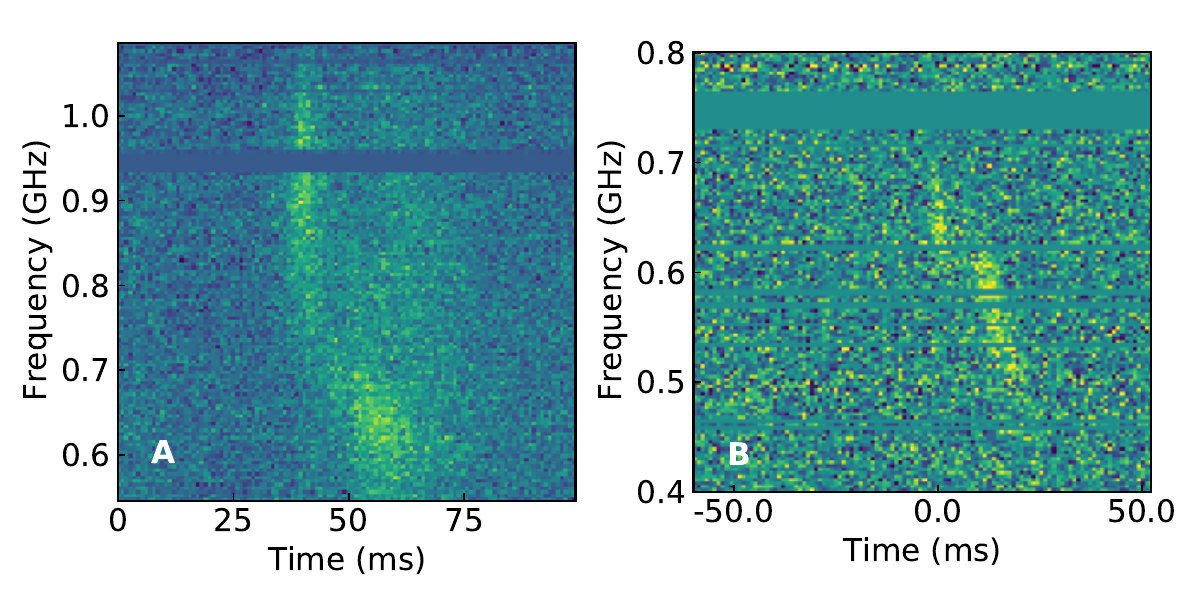}
    \caption{\textbf{Comparison of spectra of the down-drifting structures between {\GPM} and FRB\,20201229C}. (A) Spectrum of the down-drifting sub-structure within the P3 pulse of {\GPM}. (B) Spectrum of the repeating source FRB\,20201229C \cite{CHIME2023ApJ}.}
    \label{fig:down-drifting-compare}
\end{figure}

\clearpage

\begin{figure}[h!] 
    \centering
    \includegraphics[width=0.8\textwidth]{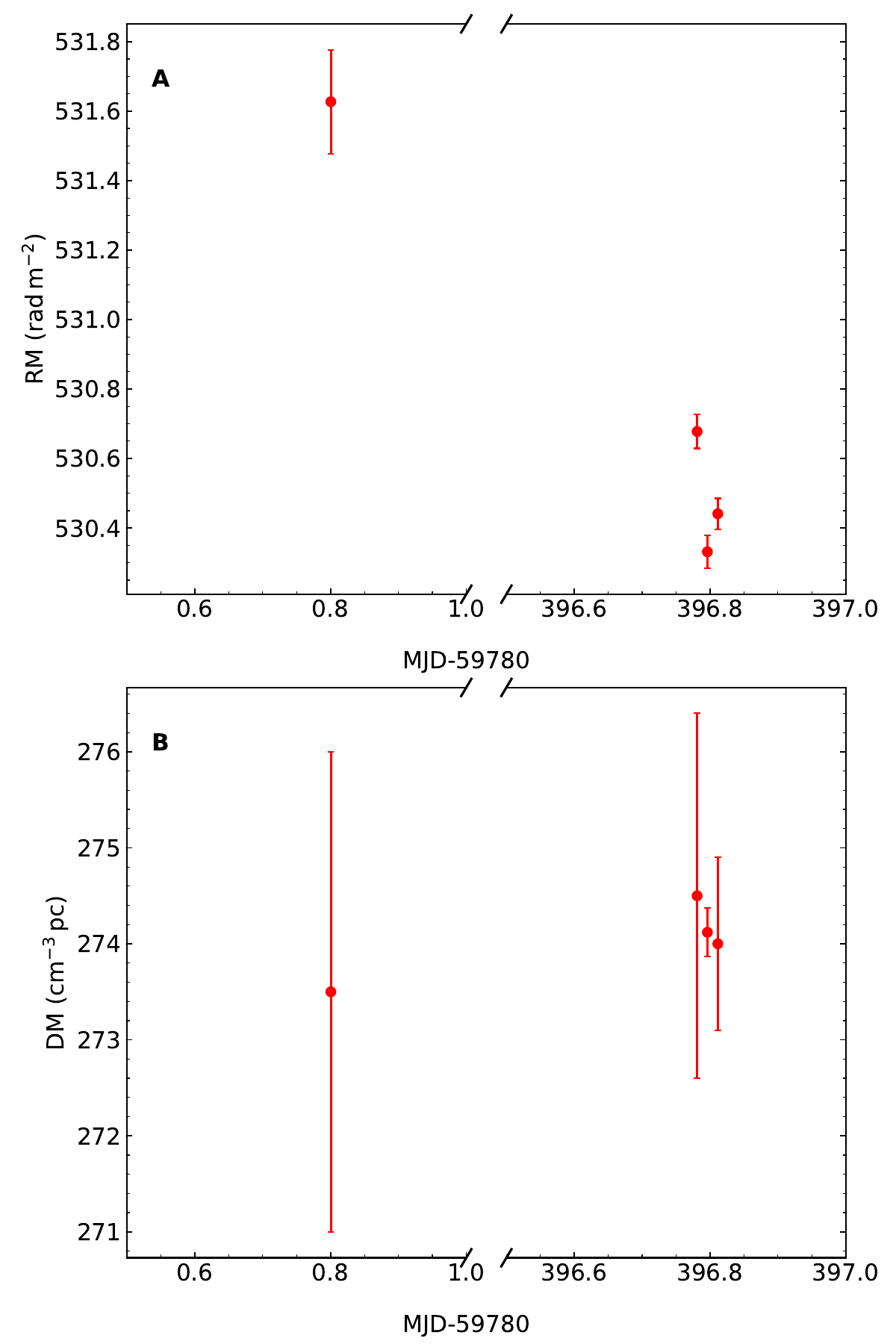}
    \caption{\textbf{Variations in RMs and DMs of {\GPM} observed on July 20, 2022\,(UT) and August 20, 2023\,(UT).} (A) RM variations with correction for the ionosphere's RM contribution (see materials and methods). (B) DM variations.}
    \label{fig:rm_dm_change}
\end{figure}

\clearpage

\begin{figure}[h!] 
    \centering
    \includegraphics[width=0.8\textwidth]{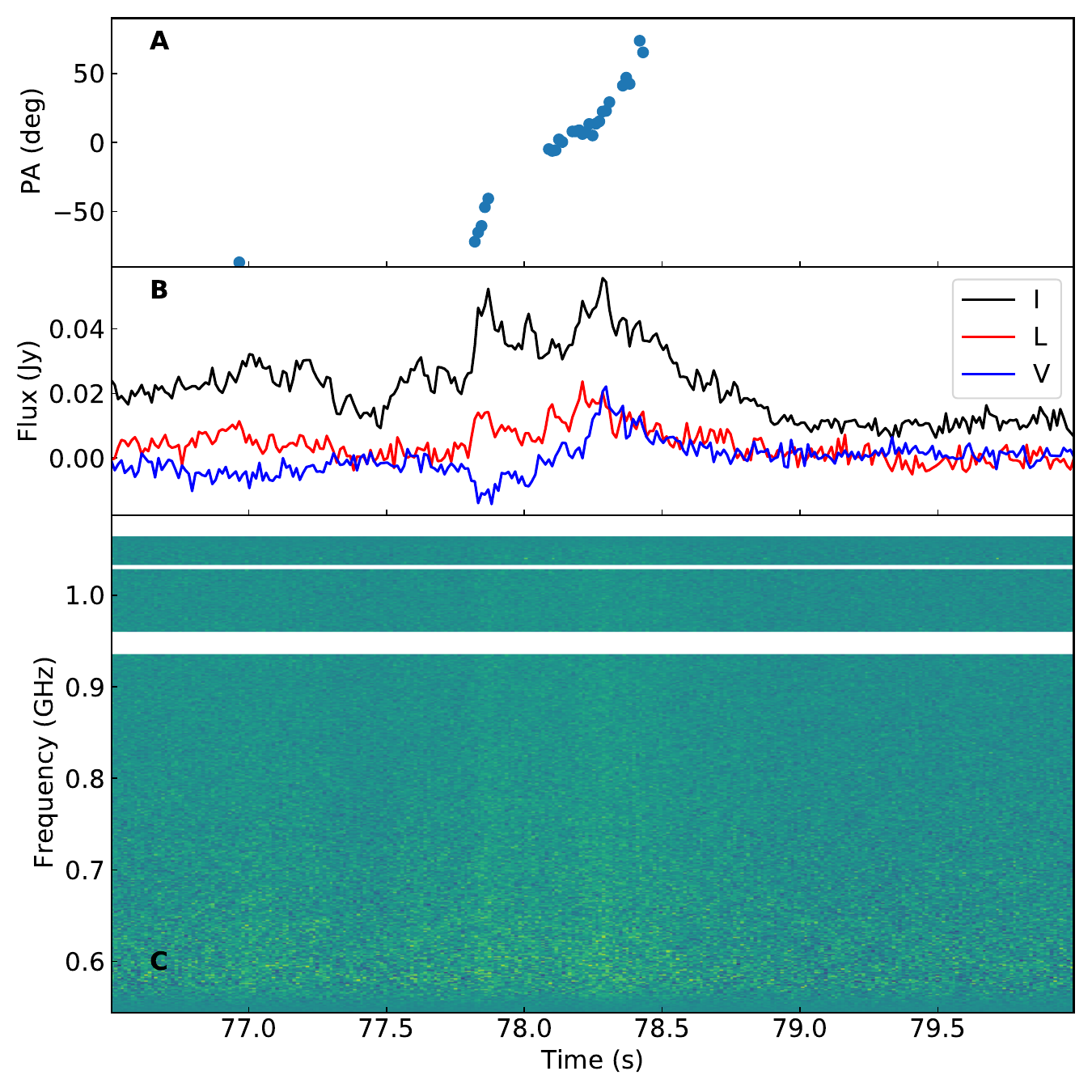}
    \caption{\textbf{Dynamic spectra of a sub-pulse exhibiting sign change in circular polarization coinciding with rapid PA change.} The sub-pulse is in period P3. (A) PA curve. (B) Total intensity (black), linear polarization intensity (red) and circular polarization intensity (blue). (C) Dynamic spectrum.}
    \label{fig:sign_change}
\end{figure}

\clearpage

\begin{figure}[h!] 
    \centering
    \includegraphics[width=0.6\textwidth]{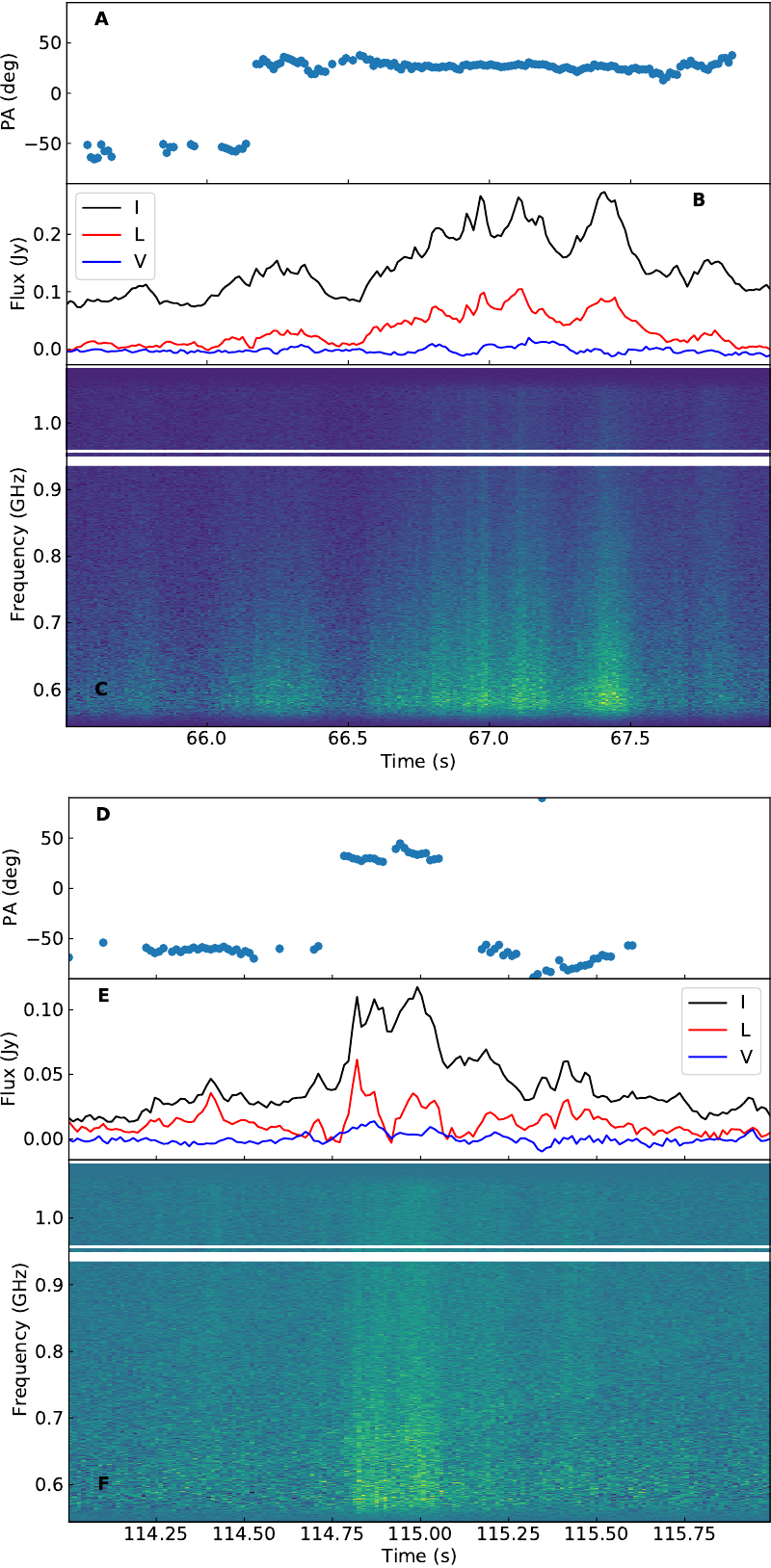}
    \caption{\textbf{Dynamic spectra of sub-pulses with orthogonal jumps.} Both sub-pulses are in period P2. (A) (D) PA curve. (B) (E) Total intensity (black), linear polarization intensity (red) and circular polarization intensity (blue). (C) (F) Dynamic spectra.}
    \label{fig:opm}
\end{figure}

\clearpage

\begin{figure}[h!] 
    \centering
    \includegraphics[width=\textwidth]{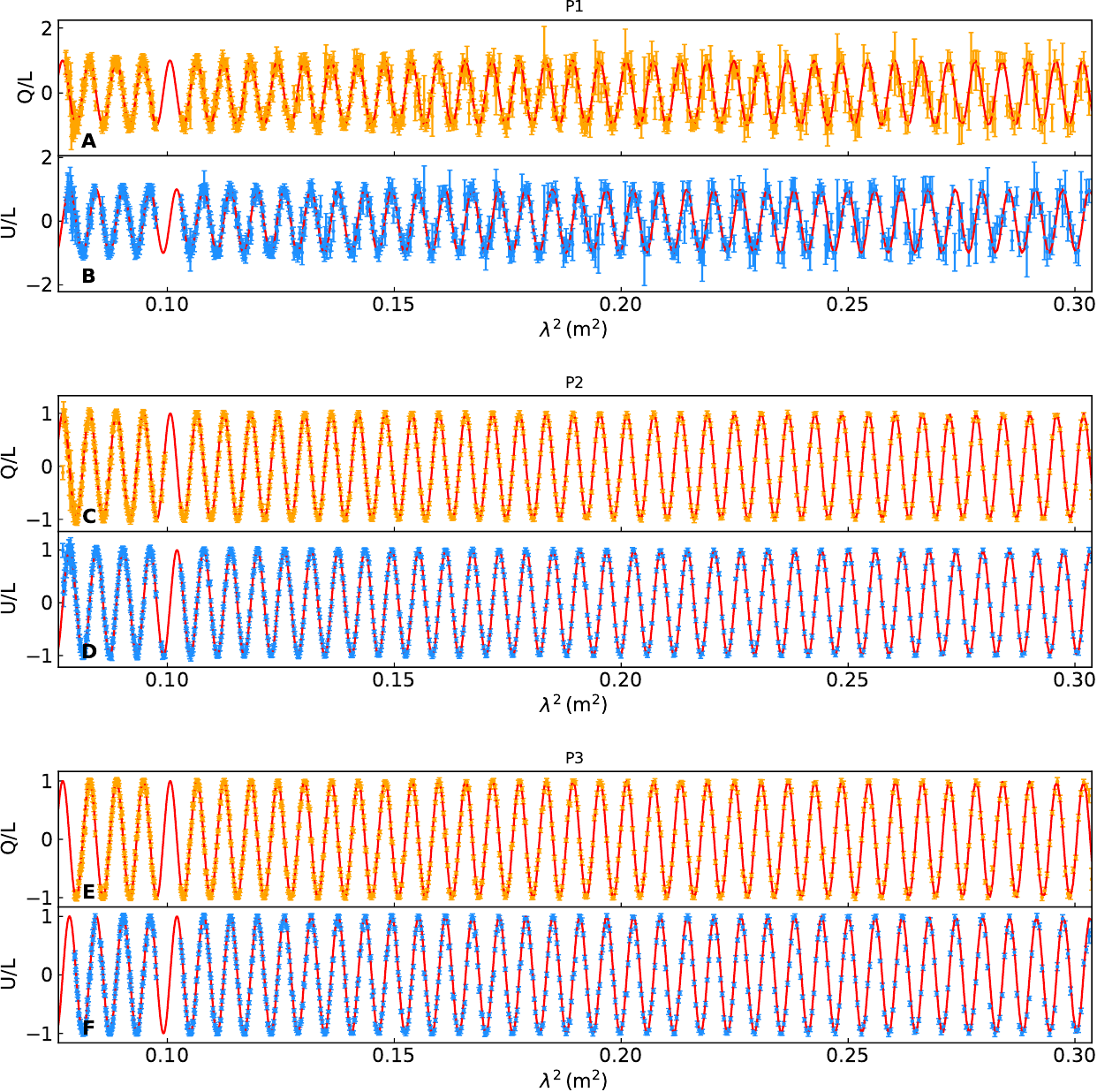}
    \caption{\textbf{Fitted curves for Stokes Q and U as a function of the square of the wavelength derived from fitting the QU spectrum.} (A) (C) (E) Fitting for the normalized Stokes Q intensity. (B) (D) (F) Fitting for the normalized Stokes U intensity.}
    \label{fig:qu_fitting}
\end{figure}

\clearpage

\begin{figure}[h!]
    \centering
    \includegraphics[width=\textwidth]{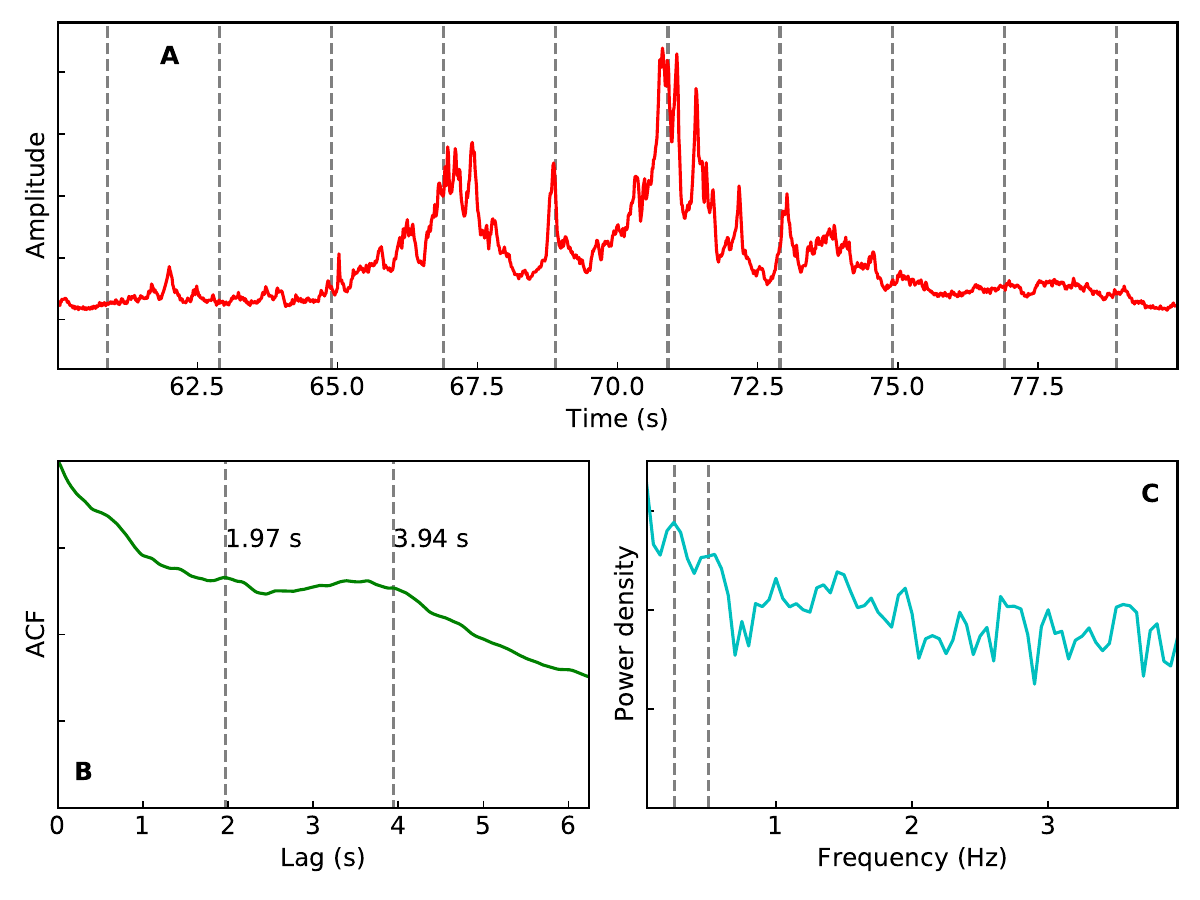}
    \caption{\textbf{Periodicity analysis of the P2 pulse.} (A) Total intensity profile of the P2 pulse within the time range of 60-80\,s. (B) Auto-correlation function (ACF). The gray dashed lines highlight the peak around 1.97\,s and its double period, 3.94\,s. (C) Power spectral density (PSD). The corresponding frequencies at 1.97\,s and 3.94\,s are marked by the gray dashed lines.}
    \label{fig:quasi-periodicity}
\end{figure}

\clearpage

\begin{figure}[h!] 
    \centering
    \includegraphics[width=\textwidth]{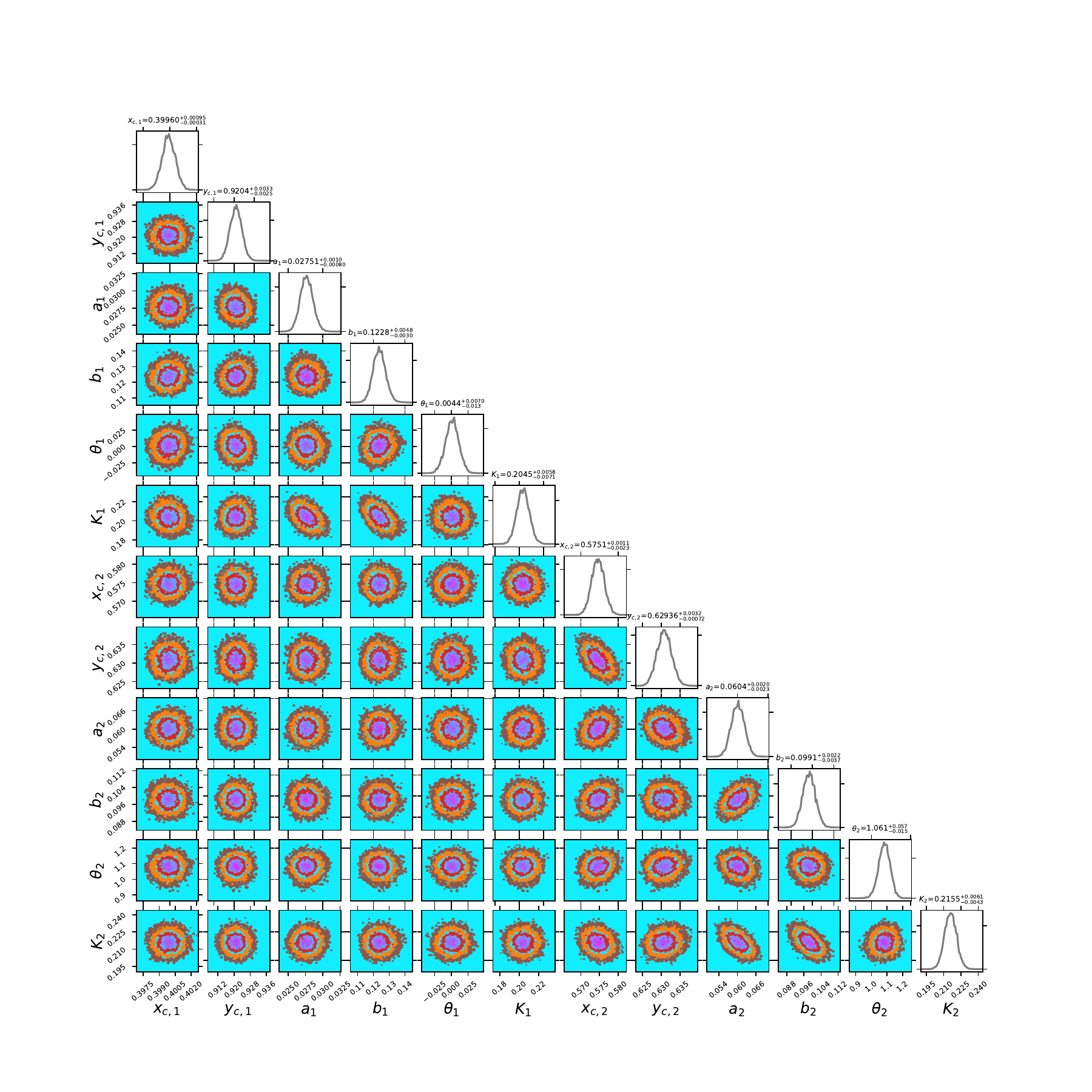}
    \caption{\textbf{Posterior distributions illustrating the parameter estimation from spectrum fitting of the down-drifting sub-structure.}}
    \label{fig:drifting_posteriors}
\end{figure}

\clearpage

\begin{figure}[h!] 
    \centering
    \includegraphics[width=0.8\textwidth]{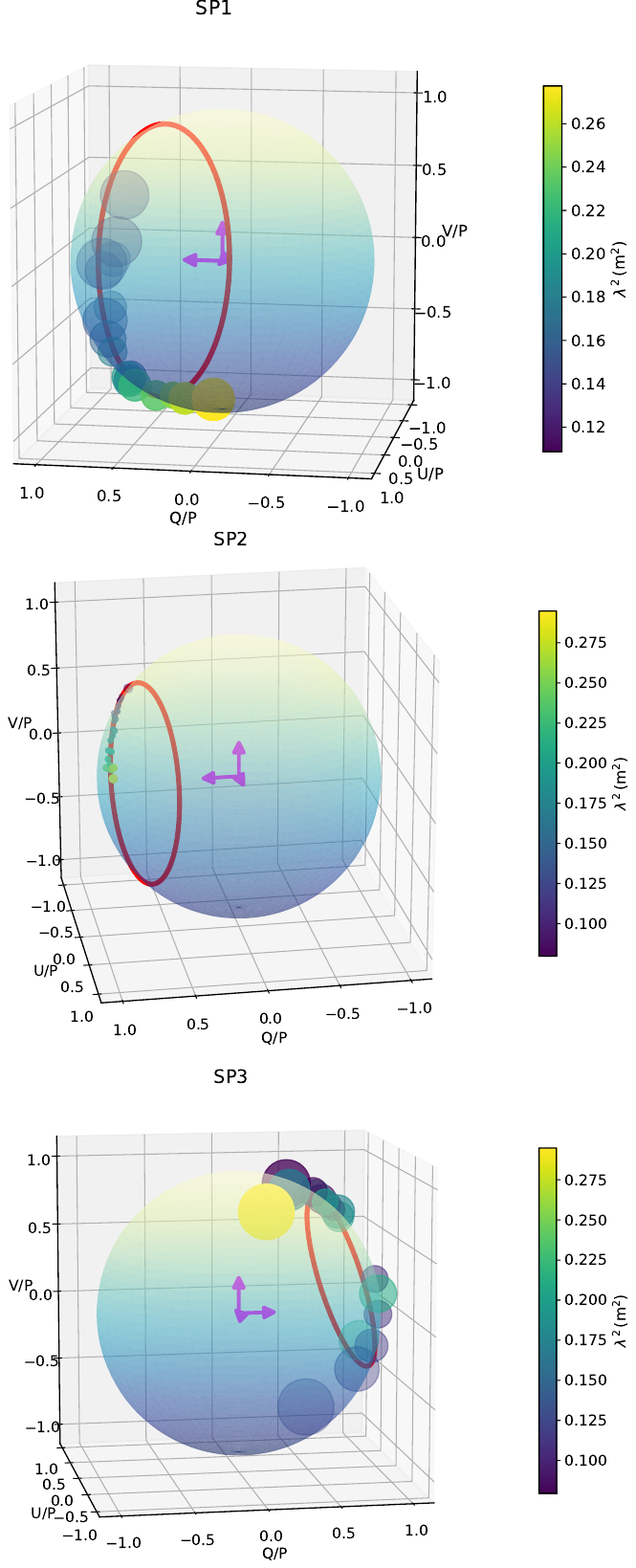}
    \caption{\textbf{The Poincaré sphere representation of the polarization vectors for the three sub-pulses, displaying linear-to-circular polarization conversion.} The red circle denotes the Faraday conversion fitting (see materials and methods). The size of each point corresponds to the error bars of the polarization vectors on the Poincaré sphere.}
    \label{fig:poincare}
\end{figure}

\clearpage

\begin{figure}[h!] 
    \centering
    \includegraphics[width=\textwidth]{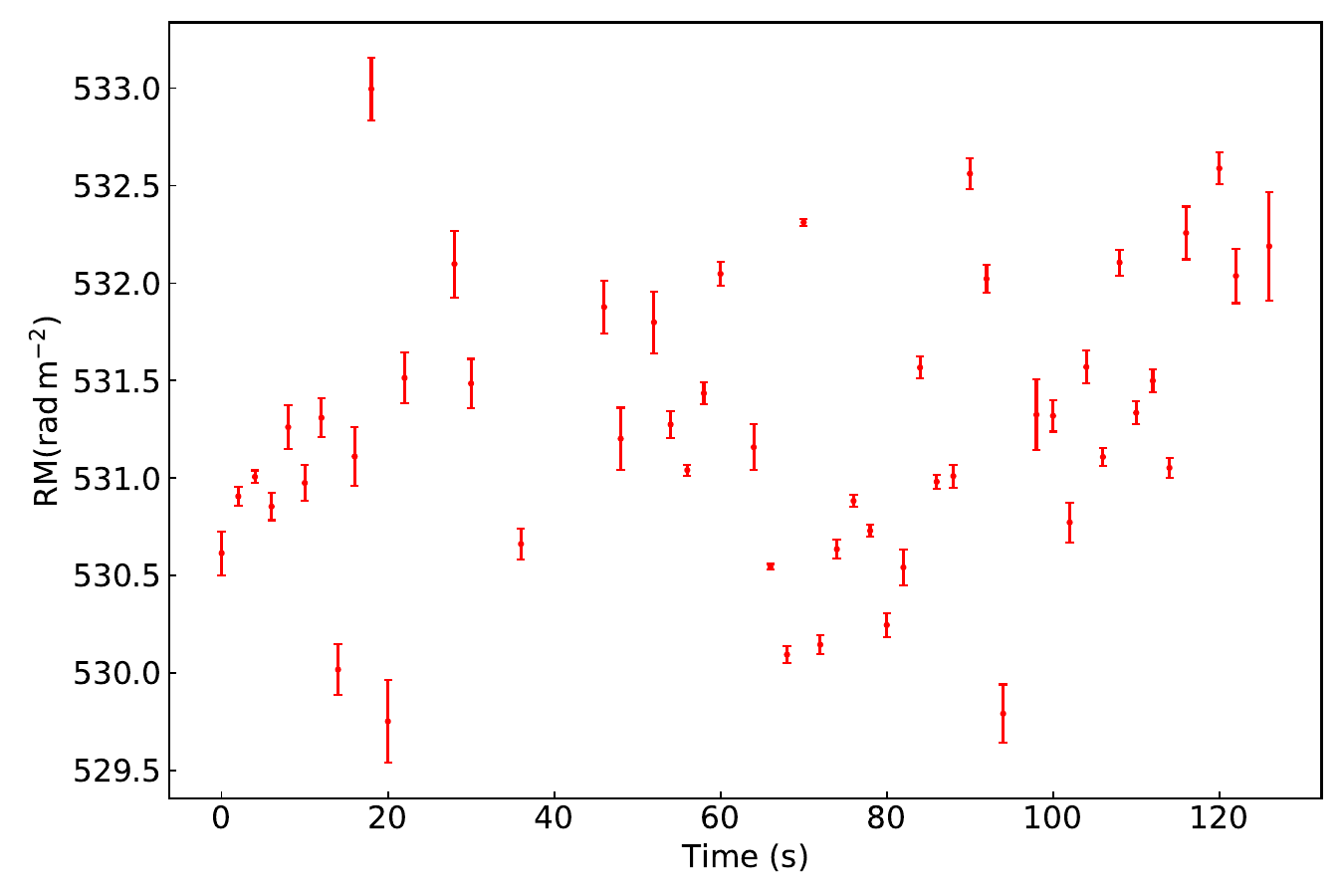}
    \caption{\textbf{Time-resolved RM variation.} The RM measurement was conducted for each two-second data block. Data points displaying an RM deviation exceeding 2\,{\radm} from the baseline of 531\,{\radm} were removed due to the weak linear polarized intensity.}
    \label{fig:RM_variation2}
\end{figure}

\clearpage

\begin{figure}[h!] 
    \centering
    \includegraphics[width=\textwidth]{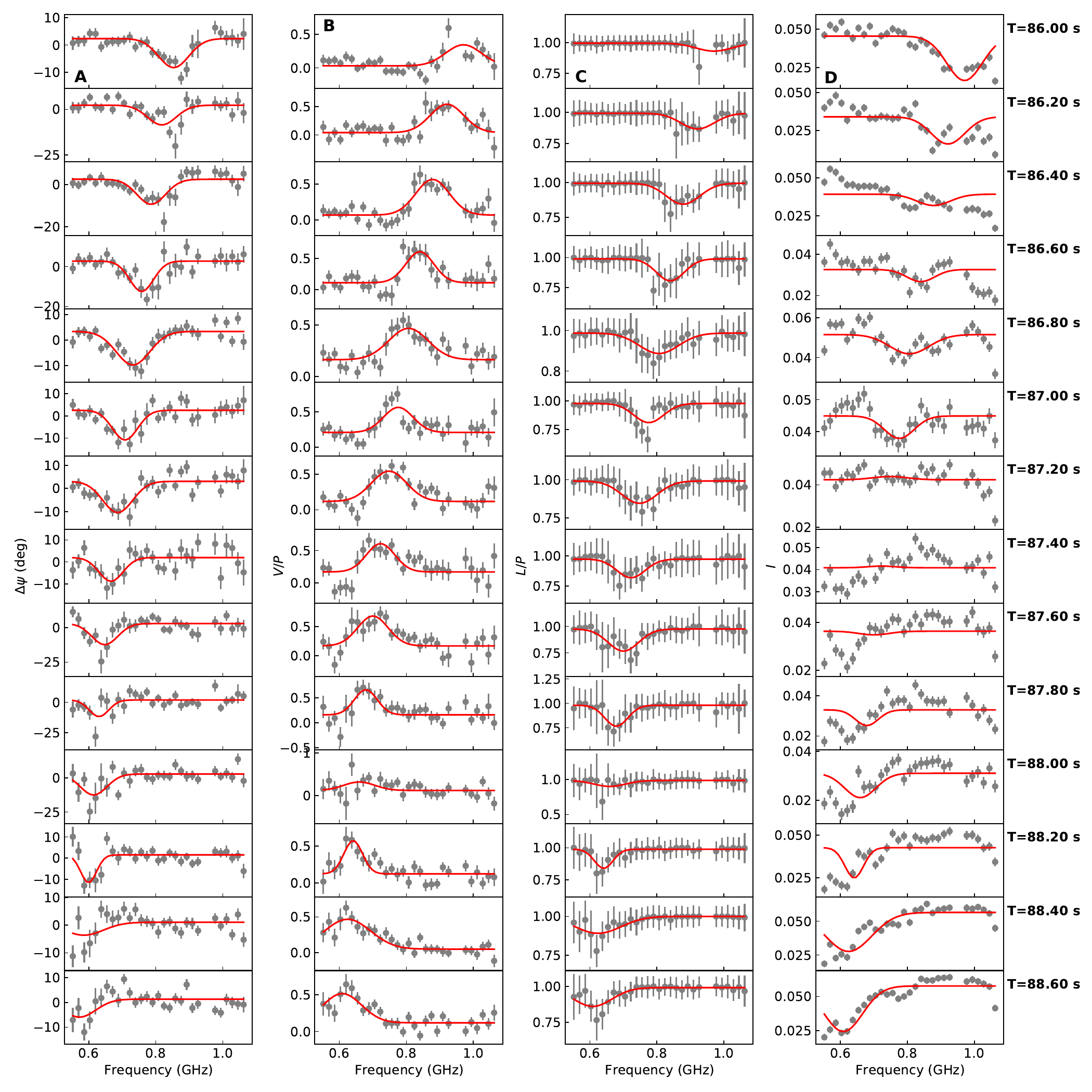}
    \caption{\textbf{Frequency-dependent PA and CP/LP variations for successive 0.2-second data segments.} (A) Fitting of PA variation. (B) Fitting of frequency-dependent circular polarization variation. (C) Fitting of frequency-dependent linear polarization variation. (D) Fitting of frequency-dependent total intensity variation. The red solid lines represent the fitted Gaussian shapes with a frequency-dependent relation (see materials and methods).}
    \label{fig:drift_conversion_fitting}
\end{figure}

\clearpage

\begin{table}[!htp]
\centering
\caption{\textbf{Follow-up observation schedule of {\GPM}.} The columns provide the start time of the session, the central observing frequency $f$,
bandwidth $\Delta f$, integration time $t_\mathrm{obs}$ and time resolution $\delta t$ of the recorded data. Sessions with detected pulsed radio emission are indicated by an asterisk.}
\label{tab:observation}

\begin{tabular}{cccccc}
\hline
\hline
Session & Start Time & $f$ (MHz)  & $\Delta f$ (MHz) & $t_{\rm obs}$ (min) & $\delta t$ ($\mu$s)\\
\hline
 1 & 2023-08-20 14:17:02 & 816 & 544 & 10 & 15 \\
 2 & 2023-08-20 14:39:02 & 816 & 544 & 10 & 15 \\
 3 & 2023-08-20 15:01:03 & 816 & 544 & 10 & 15 \\
 4 & 2023-08-20 15:23:03 & 816 & 544 & 10 & 15 \\
 5 & 2023-08-20 15:45:05 & 816 & 544 & 10 & 15 \\
 6 & 2023-08-20 16:07:06 & 816 & 544 & 10 & 15 \\
 7 & 2023-08-20 16:29:07 & 816 & 544 & 10 & 15 \\
 8 & 2023-08-20 16:56:03 & 816 & 544 & 10 & 15 \\
 9 & 2023-08-20 17:15:31 & 816 & 544 & 10 & 15 \\
 10 & 2023-08-20 17:37:36 & 816 & 544 & 10 & 15 \\
 11 & 2023-08-20 17:59:43 & 816 & 544 & 10 & 15 \\
 12 & 2023-08-20 18:21:53 & 816 & 544 & 10 & 15 \\
 13$^{\ast}$ & 2023-08-20 18:44:02 & 816 & 544 & 10 & 15 \\
 14$^{\ast}$ & 2023-08-20 19:06:12 & 816 & 544 & 10 & 15 \\
 15$^{\ast}$ & 2023-08-20 19:28:22 & 816 & 544 & 10 & 15 \\
\hline
\end{tabular}
\end{table}
\clearpage

\end{document}